\documentclass[useAMS,usegraphicx,usenatbib,letterpaper]{emulateapj}

\usepackage{ulem}
\usepackage{color}
\usepackage{graphics,graphicx}
\usepackage{times} 
\usepackage{amssymb}
\usepackage{amsmath}
\usepackage{natbib}
\usepackage{url}
\usepackage{multirow}
\usepackage{ctable}
\usepackage{threeparttable}
\newif\ifAMStwofonts
\AMStwofontstrue

\def\xmm{{\it XMM-Newton}}

\def\suzaku{{\it Suzaku}}

\def\chandra{{\it Chandra}}
\def\swift{{\it Swift}}

\def\epicpn{{\it EPIC}{\rm-pn}}
\def\epicmos1{{\it EPIC}{\rm-MOS1~\/}}
\def\epicmos2{{\it EPIC}{\rm-MOS2 ~\/}}
\def\epicmos{{\it EPIC}{\rm-MOS}}

\def\nustar{{\it NuSTAR}}
\def\spitzer{{\it Spitzer}}

\def\deg{$^{\circ}$}

\def\H0{{\rm ~km~s^{-1}~Mpc^{-1}}}

\def\kev{\hbox{\rm keV}}

\def\atpcm{{\rm atom~cm$^{-2}$}}

\def\ergpcmsqps{\hbox{$\rm\thinspace erg~cm^{-2}~s^{-1}$}}

\def\ergps{\hbox{erg~s$^{-1}$}}

\def\msun{\hbox{$\rm\thinspace M_{\odot}$}}

\def\chisq{{$\chi^{2}$}}
\def\dchisq{$\Delta\chi^{2}$}
\def\rchi{{$\chi^{2}_{\nu}$}}

\def\xspec{\hbox{\small XSPEC}}

\def\heasoft{\hbox{\rm{\small HEASOFT}}}
\def\xselect{\hbox{\rm{\small XSELECT}}}
\def\sas{\hbox{\rm{\small SAS}}}
\def\xmmselect{\hbox{\rm{\small XMMSELECT}}}
\def\epchain{\hbox{\rm{\small EPCHAIN}}}
\def\emchain{\hbox{\rm{\small EMCHAIN}}}

\def\rmfgen{\hbox{\rm{\small RMFGEN}}}
\def\arfgen{\hbox{\rm{\small ARFGEN}}}

\def\epiclccorr{\hbox{\rm{\small EPICLCCORR}}}

\def\nustardas{\rm{\small NUSTARDAS}}
\def\xrtpipeline{\rm{\small XRTPIPELINE}}
\def\xrtmkarf{\rm{\small XRTMKARF}}
\def\ftool{\hbox{\rm{\small FTOOL}}}

\def\grppha{\hbox{\rm{\small GRPPHA}}}

\def\addascaspec{\hbox{\rm{\small ADDASCASPEC}}}

\def\grid25{\hbox{\rm{\small GRID25}}}
\def\pl{\rm{\small POWERLAW}}

\def\tbabs{\rm{\small TBABS}}
\def\tbnew{\rm{\small TBNEW}}
\def\tbnewlink{http://pulsar.sternwarte.uni-erlangen.de/wilms/research/tbabs}
\def\diskbb{\rm{\small DISKBB}}
\def\diskpbb{\rm{\small DISKPBB}}

\def\cflux{\rm{\small CFLUX}}

\def\reflionx{\rm{\small REFLIONX}}

\def\relconv{\rm{\small RELCONV}}
\def\kerrbb{\rm{\small KERRBB}}

\def\comptt{\rm{\small COMPTT}}

\def\simpl{\rm{\small SIMPL}}

\def\ka{K$\alpha$}

\def\feabund{$A_{\rm Fe}$}
\def\zsun{$Z_{\rm \odot}$}

\def\etal{et al.}
\def\eg{{\it e.g.~\/}}
\def\etc{{\it etc.}}
\def\ie{{\it i.e.~\/}}

\def\la{\mathrel{\hbox{\rlap{\hbox{\lower4pt\hbox{$\sim$}}}{\raise2pt\hbox{$<$}}}}}
\def\ga{\mathrel{\hbox{\rlap{\hbox{\lower4pt\hbox{$\sim$}}}{\raise2pt\hbox{$>$}}}}}

\def\d25{D$_{25}$}

\def\.25{0.25 keV\thinspace}
\def\lx{$L_{\rm X}$}

\def\mbh{\rm $M_{\rm BH}$}

\def\rin{$R_{\rm in}$}

\def\Tin{$T_{\rm in}$}
\def\Teff{$T_{\rm eff}$}
\def\fcol{$f_{\rm col}$}

\def\fvar{$F_{\rm var}$}

\def\culx{Circinus ULX5}
\def\hoix{Holmberg\,IX X-1}

\shorttitle{An Extremely Luminous, Variable ULX in Circinus}
\shortauthors{D.~J. Walton, et al.}

\begin{document}

\title{An Extremely Luminous and Variable Ultraluminous X-ray Source in the Outskirts of Circinus Observed with \nustar}

\author{D. J. Walton\altaffilmark{1},
F. Fuerst\altaffilmark{1},
F. A. Harrison\altaffilmark{1},
D. Stern\altaffilmark{1,2},
M. Bachetti\altaffilmark{3,4},
D. Barret\altaffilmark{3,4},
F. E. Bauer\altaffilmark{5,6},
S. E. Boggs\altaffilmark{7},
F. E. Christensen\altaffilmark{8},
W. W. Craig\altaffilmark{7,9},
A. C. Fabian\altaffilmark{10},
B. W. Grefenstette\altaffilmark{1},
C .J. Hailey\altaffilmark{11},
K. K. Madsen\altaffilmark{1},
J. M. Miller\altaffilmark{12},
A. Ptak\altaffilmark{13},
V. Rana\altaffilmark{1},
N. A. Webb\altaffilmark{3,4},
W. W. Zhang\altaffilmark{13}
}
\affil{
$^{1}$Space Radiation Laboratory, California Institute of Technology, Pasadena,
CA 91125, USA \\
$^{2}$Jet Propulsion Laboratory, California Institute of Technology, Pasadena, CA 91109, USA \\
$^{3}$Universite de Toulouse; UPS-OMP; IRAP; Toulouse, France \\
$^{4}$CNRS; IRAP; 9 Av. colonel Roche, BP 44346, F-31028 Toulouse
cedex 4, France \\
$^{5}$Instituto de Astrof\'{\i}sica, Facultad de F\'{i}sica, Pontificia Universidad Cat\'{o}lica de Chile, 306, Santiago 22, Chile \\
$^{6}$Space Science Institute, 4750 Walnut Street, Suite 205, Boulder, Colorado 80301 \\
$^{7}$Space Sciences Laboratory, University of California, Berkeley, CA 94720, USA \\
$^{8}$DTU Space, National Space Institute, Technical University of Denmark, Elektrovej 327, DK-2800 Lyngby, Denmark \\
$^{9}$Lawrence Livermore National Laboratory, Livermore, CA 94550, USA \\
$^{10}$Institute of Astronomy, University of Cambridge, Madingley
Road, Cambridge CB3 0HA, UK \\
$^{11}$Columbia Astrophysics Laboratory, Columbia University, New York, NY 10027, USA \\
$^{12}$Department of Astronomy, University of Michigan, 500
Church Street, Ann Arbor, MI 48109-1042, USA \\
$^{13}$NASA Goddard Space Flight Center, Greenbelt, MD 20771,
USA \\
}

\begin{abstract}
Following a serendipitous detection with the \nustar\ observatory, we present
a multi-epoch spectral and temporal analysis of an extremely bright
ultraluminous X-ray source (ULX) located in the outskirts of the Circinus galaxy
($\sim$4$'$ away from the nucleus), hereafter \culx, including coordinated
follow-up observations with \xmm\ and \nustar. The \nustar\ data presented
here represent one of the first instances of a ULX reliably detected at hard
($E>10$\,\kev) X-rays. \culx\ is variable on long timescales by at least a factor
of  $\sim$5 in flux, and was caught in a historically bright state during our 2013
observations, with an observed 0.3--30.0\,\kev\ luminosity of
$1.6\times10^{40}$\,\ergps. During this epoch, the source displayed a curved
3--10\,\kev\ spectrum, broadly similar to other bright ULXs. Although pure
thermal models result in a high energy excess in the \nustar\ data, this excess
is too weak to be modelled with the disk reflection interpretation previously
proposed to explain the 3--10\,\kev\ curvature in other ULXs. In addition to flux
variability, \culx\ also displays clear spectral variability. While in many cases the
interpretation of spectral components in ULXs is uncertain, the spectral and
temporal properties of the all the high quality datasets currently available
strongly support a simple disk--corona model reminiscent of that invoked for
Galactic binaries, with the accretion disk becoming more prominent as the
luminosity increases. However, although the disk temperature and luminosity are
remarkably well correlated across all timescales currently probed, the observed
relation is $L\propto T^{1.70\pm0.17}$, flatter than that expected for simple
blackbody radiation. The spectral variability displayed by \culx\ is highly
reminiscent of that observed from the Galactic black hole binaries (BHBs)
XTE\,J1550-564 and GRO\,J1655-40 at high luminosities. This comparison
would  imply a black hole mass of $\sim$90\ \msun\ for \culx. However, given
the diverse behavior observed from Galactic BHB accretion disks, this mass
estimate is still uncertain. Finally, during our coordinated \xmm+\nustar\
observation we find no evidence for any ionised iron absorption lines, typically
associated with disk winds in Galactic BHBs. The limits placed on any undetected 
features imply that we are not viewing the central regions of \culx\ through any
extreme super-Eddington outflow.
\end{abstract}


\section{Introduction}

The origin of the extreme luminosities displayed by ultraluminous X-ray sources
(ULXs; \lx\,$\gtrsim$\,$10^{39}$\,\ergps) may relate to exotic super-Eddington
modes of accretion (\eg \citealt{Poutanen07}, \citealt{Finke07}), or alternatively
to the presence of black holes larger than typically found in Galactic black hole
binary systems (BHBs; \mbh\ $\sim$10\msun), potentially the long sought
intermediate mass black holes (IMBHs: $10^{2}$ $\lesssim$ \mbh\ $\lesssim$
$10^{5}$\,\msun; \eg \citealt{Miller04, Strohmayer09b}). For recent reviews
see \cite{Roberts07rev} and \cite{Feng11rev}.

The majority of ULXs only modestly exceed $10^{39}$\,\ergps, and therefore
likely represent a natural extension of the disk--dominated high-Eddington
thermal states displayed by Galactic BHBs (\eg \citealt{Kajava09,
Middleton13nat}). However, a much smaller fraction of the ULX population have
been observed to exceed $10^{40}$\,\ergps\ in X-rays (\citealt{WaltonULXcat,
Jonker12, Sutton12}), and in the most luminous case known to date to reach as
high as $\sim$10$^{42}$\,\ergps\ (\citealt{Farrell09}). These more luminous
sources, apparently radiating in excess of 10 times the Eddington limit for a
10\,\msun\ black hole, remain among the best known candidates to host
massive black holes.

Here we report on an extremely bright and highly variable ULX in the outskirts
of the Circinus galaxy ($z$ = 0.001448, $D \sim 4$\,Mpc; \citealt{Freeman77,
Koribalski04}), hereafter \culx\ (there are up to four other known/claimed ULX
candidates closer to the Circinus nucleus, see \eg \citealt{Bauer01, Swartz04,
LiuMirabel05, Ptak06}). \culx\ was serendipitously detected in a high luminosity
state by \nustar\ on 25th Jan 2013, and we subsequently performed follow-up
target-of-opportunity (ToO) observations with both \xmm\ and the recently
launched Nuclear Spectroscopic Telescope Array (\nustar; \citealt{NUSTAR}).
In spite of its luminosity, \culx\ has received little attention to date, owing in
large part to its fairly substantial separation from the galaxy centre (\culx\
formally falls outside the D25 isophote for Circinus). The only mention of the
source is in the \cite{Winter06} ULX catalogue.\footnote{Note that in
\cite{Winter06}, the source studied here is referred to as Circinus XMM2.} In 
section \ref{sec_red} we detail our data reduction procedure for the various
datasets considered, and in sections \ref{sec_spec}, \ref{sec_shortvar} and 
\ref{sec_longvar} we describe our multi-epoch spectral and temporal analysis of
this remarkable source. Key results are discussed in section \ref{sec_dis}, and we
summarise our conclusions in section \ref{sec_conc}. Throughout this work, we
will assume \culx\ is indeed associated with the Circinus galaxy. Possible
alternative scenarios are discussed in section \ref{sec_circinus}, but we consider
them highly unlikely.

\section{Data Reduction}
\label{sec_red}

Here we outline our data reduction procedure for the X-ray observations
considered in this work, beginning with the new \xmm\ and \nustar\ datasets
obtained in early 2013.

\begin{table*}
  \caption{Details of the X-ray observations considered in this work, ordered
  chronologically.}
  \begin{center}
\begin{tabular}{c c c c c c}
\hline
\hline
\\[-0.25cm]
Mission & OBSID & Date & Target & Good Exposure & ULX5 3-10\,\kev\ Flux\tmark[b] \\
& & & & (ks) & ($10^{-12}$ \ergpcmsqps) \\
\\[-0.3cm]
\hline
\hline
\\[-0.2cm]
\xmm\ & 0111240101 & 2001-08-06 & Nucleus & 105/110/-\tmark[a] &  $1.09 \pm 0.04$ \\
\\[-0.225cm]
\suzaku\ & 701036010 & 2006-07-21 & Nucleus & 110 & $1.55 \pm 0.04$ \\
\\[-0.225cm]
\swift\ & 00035876001 & 2007-03-23 & Field & 7.5 & $0.8^{+0.1}_{-0.3}$ \\
\\[-0.225cm]
\swift\ & 00037273001 & 2008-05-18 & Field & 6 & $1.0^{+0.3}_{-0.4}$ \\
\\[-0.225cm]
\chandra\ & 10873 & 2009-03-01 & SN1996cr & 18 & $5.0 \pm 0.5$ \\
\\[-0.225cm]
\chandra\ & 10850 & 2009-03-03 & SN1996cr & 14 & $4.8^{+0.4}_{-0.5}$ \\
\\[-0.225cm]
\chandra\ & 10872 & 2009-03-04 & SN1996cr & 17 & $1.0^{+0.2}_{-0.1}$ \\
\\[-0.225cm]
\swift & 00090260001 & 2009-11-15 & Field & 5 & $2.3^{+1.0}_{-0.6}$ \\
\\[-0.225cm]
\swift & 00037273004 & 2012-11-11 & Field & 3.5 & $1.0^{+0.7}_{-0.5}$ \\
\\[-0.225cm]
\nustar\ & 60002039002 & 2013-01-25 & Nucleus & 55 & $4.2 \pm 0.2$ \\
\\[-0.225cm]
\swift\ & 00032699001 & 2013-01-31 & Field & 3.5 & $5.0^{+1.3}_{-1.0}$ \\
\\[-0.225cm]
\nustar\ & 30002038002 & 2013-02-02 & ULX5 & 18 & $4.2 \pm 0.2$ \\
\\[-0.225cm]
\xmm\ & 0701981001 & 2013-02-03 & ULX5 & 37/47/47\tmark[a] & (as below) \\
\\[-0.225cm]
\nustar & 30002038004 & 2013-02-03 & ULX5 & 40 & $4.93\pm0.12$ \\
\\[-0.225cm]
\nustar & 30002038006 & 2013-02-05 & ULX5 & 36 & $3.7 \pm 0.1$ \\
\\[-0.2cm]
\hline
\hline
\\[-0.15cm]
\end{tabular}
\\
$^{a}$ \xmm\ exposures are listed for the \epicpn/MOS1/MOS2 detectors. \\
$^{b}$ Observed fluxes are computed in sections \ref{sec_spec} for the higher
S/N observations, utilizing the \diskbb+\simpl\ model, and in \ref{sec_snapshot}
for the shorter shapshot observations, utilizing a simpler cutoff-powerlaw model.
\vspace*{0.3cm}
\label{tab_obs}
\end{center}
\end{table*}

\subsection{2013 Observations}

\subsubsection{NuSTAR}

\begin{figure}
\hspace*{-0.2cm}
\epsscale{1.15}
\rotatebox{0}{
\plotone{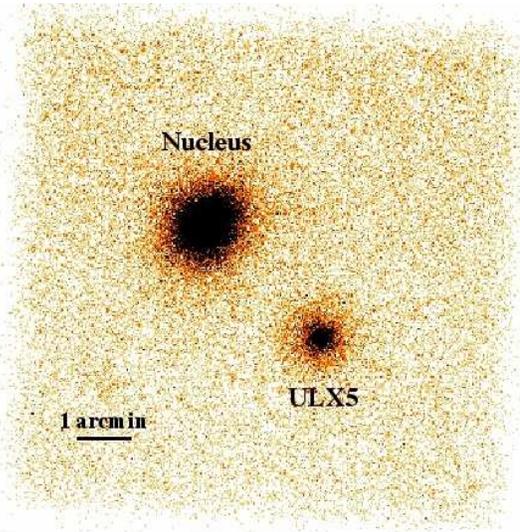}}
\caption{
\nustar\ FPMA image of the Circinus field. The Circinus nucleus and \culx,
separated by $\sim$4', are both highlighted.}
\label{fig_ulx}
\vspace*{0.2cm}
\end{figure}

\nustar\ (\citealt{NUSTAR}) performed four observations of the Circinus field
throughout late January and early February 2013. The first observation, in which
the bright ULX was serendipitously detected, was optimized to observe the
galaxy's nucleus, but the three subsequent follow-up observations were all
optimized to observe the ULX, which is offset from the nucleus by $\sim$4' (see
Fig. \ref{fig_ulx}): RA = $14^{h} 12^{m} 39^{s}$, DEC = $-65$\deg $23' 34''$.
The data have been reduced using the standard pipeline, part of the \nustar\
Data Analysis Software v0.11.1 (\nustardas, part of the standard \heasoft\
distribution as of version 14), and instrumental responses from \nustar\
caldb\footnote{The \nustar\ calibration database is available through HEASARC:
http://heasarc.gsfc.nasa.gov/docs/heasarc/caldb/nustar} v20130509 are used
throughout. As discussed in \cite{Risaliti13nat}, these responses have been
empirically corrected such that the Crab nebula gives a powerlaw spectrum with
a photon index of $\Gamma_{\rm Crab} = 2.1$. Full details are provided in
Madsen et al. 2013 (\textit{in prep.}), but the calibration has also been
successfully tested against other powerlaw-like sources, including
the pulsar wind nebula G\,21.5-0.9 and the blazar PKS\,2155, which are
frequently used to assess the calibration of X-ray missions (\citealt{Tsujimoto11,
Ishida11}). The unfiltered event files have been cleaned with the standard depth
correction, which significantly reduces the internal background at high energies,
and SAA passages have been removed. Source products were obtained from
circular regions of $\sim$70$''$ in radius for the observations with the ULX on
the optical axis, and $\sim$50$''$ in radius for the initial detection to avoid the
edge of the detector, and background was estimated from a blank area of the
same detector free from contaminating point sources. Spectra and lightcurves
were extracted from the cleaned event files using \xselect\ for both focal plane
modules (FPMA and FPMB). Finally, the spectra were grouped with \grppha\ such
that each spectral bin contains at least 50 counts. Although \nustar\ operates
over the 3-79\ \kev\ energy range, \culx\ is only reliably detected up to
$\sim$30\,\kev.

\subsubsection{XMM-Newton}
\label{sec_red_xmm1}

After the initial serendipitous \nustar\ detection, we also triggered a
$\sim$50\,ks target-of-opportunity (ToO) observation with \xmm\
(\citealt{XMM}) in order to  provide complimentary soft X-ray coverage for our
targeted \nustar\ follow-up observations (performed simultaneously with
\nustar\ obsid 30002038004). Data reduction was carried out with the \xmm\
Science Analysis System (\sas v12.0.1) largely according to the standard
prescription provided in the online guide.\footnote{http://xmm.esac.esa.int/}
The observation data files were processed using \epchain\ and \emchain\ to
produce calibrated event lists for the \epicpn\ (\citealt{XMM_PN}) and \epicmos\ 
(\citealt{XMM_MOS}) detectors, respectively. For each detector, source products
were extracted from a circular region of $\sim$40'' in radius, and background
was estimated from an area of the same CCD free of contaminating point
sources. Lightcurves and spectra were generated with \xmmselect, selecting
only single and double events (single to quadruple events) for \epicpn\
(\epicmos), excluding periods of high background flares (occuring
predominantly at the end of the observation). The redistribution matrices and
auxiliary response files were generated with \rmfgen\ and \arfgen, while
lightcurves were corrected for the background count rate using \epiclccorr.
After performing the data reduction separately for each of the MOS CCDs and
confirming their consistency, the spectra were combined using the \ftool\
\addascaspec. Finally, spectra were re-binned to have a minimum of 50 counts
in each energy bin, and analysed across the full 0.3--10.0\,\kev\ energy range.

\pagebreak
\subsection{Archival Data}

Prompted by the new data obtained in 2013, we also searched the X-ray archive
for observations of the Circinus field, in order to investigate potential long term
variability. A summary of all the X-ray observations considered in this work is
given in Table \ref{tab_obs}. The additional archival observations were almost
all targeted at the Circinus nucleus or the nearby supernova SN1996cr
(\citealt{Bauer08}), but \culx\ is serendipitously included in the field-of-view,
albeit substantially off-axis, in each case.

\subsubsection{XMM-Newton}
\label{sec_red_xmm2}

In addition to the recent ToO we obtained, \xmm\ also observed Circinus in
August 2001. This observation was largely reduced in the same manner as the
ToO described above (section \ref{sec_red_xmm1}). However, in this case the
target was the Circinus nucleus, and the ULX in question unfortunately fell very
close to a chip gap in the \epicpn\ detector. Furthermore, the MOS2 detector
was operated in partial window mode, and the target fell outside the operational
region of the detector. In this case, we use an elliptical source region for \epicpn,
offset slightly from the source position in  order to include as many counts as
possible without the region covering the chip gap. The shape of the \epicpn\
spectrum obtained is consistent with that obtained with MOS1, however we do
find that the flux normalisation is not consistent between the detectors. We
therefore take the MOS1 detector as the true flux indicator for this observation,
given the non-standard reduction necessary for \epicpn. In this case, as the
average countrate is much lower than the more recent  \xmm\ observation,
owing to the combined effect of a large off-axis angle and a lower source flux
(see section \ref{sec_xmm1} and Table \ref{tab_obs}), we rebin instead to a
minimum of 25 counts per bin.

\pagebreak
\subsubsection{Suzaku}
\label{sec_suz_red}

The Circinus field was also observed by \suzaku\ (\citealt{SUZAKU}) in July 2006.
Owing to the obvious, dominant contribution from the nucleus, we do not
consider the HXD detectors in this work, and focus instead on the data obtained
with the XIS CCDs. Following the recommendation in the \suzaku\ Data
Reduction Guide,\footnote{http://heasarc.gsfc.nasa.gov/docs/suzaku/analysis/}
we reprocessed the unfiltered event files for each of the operational XIS detectors
(XIS0, 1, 2 and 3; \citealt{SUZAKU_XIS}) and editing modes (3x3 and 5x5) using
the latest \heasoft\ software package (v6.13). Cleaned event files were generated
by re-running the \suzaku\ pipeline with the latest calibration, as well as the
associated screening criteria files. Source products were extracted from a circular
region of $\sim$85'' in radius, in order to avoid contamination from a further
faint source nearby (separated by $\sim$105''), and the background was
extracted from regions free of any obvious contaminating point sources, but
close to the source region. Spectra and lightcurves were extracted from the
cleaned event files with \xselect, and responses were generated for each detector
using the {\small XISRESP} script with a medium resolution. The spectra and
response files for the front-illuminated detectors (XIS0, 2 and 3) were combined
using the \ftool\ \addascaspec, after confirming their consistency. Finally, we
again grouped the spectra to have a minimum of 50 counts per energy bin. In
this work, we analyse the XIS data over the 0.5--10.0\,\kev\ energy range.

\subsubsection{Chandra}

Although the Circinus field has also frequently been observed by the \chandra\
X-ray observatory (\citealt{CHANDRA}), \culx\ only fell in the field-of-view for
three of these pointings (see Table \ref{tab_obs}). For each of the three
observations, the instrument was operated in the Timed Event mode, and we
extracted spectra from the ACIS-S detector (\citealt{CHANDRA_ACIS}) using the
standard pipeline CIAO v4.5. In all observations \culx\ is detected close to the
edge of the field of view, while the observatory pointed at SN1996cr. At these
large off-axis angles the \chandra\ PSF is clearly elongated, so we used an
elliptical extraction region with major and minor axes of 5.5$\times$2.7" to
gather all source photons. The background was extracted from two large
circular regions above and below the HETG diffraction pattern, free from any
other contaminating sources. The \chandra\ spectra were rebinned to a
signal-to-noise ratio (S/N) of 2, and modelled over the full 0.3--9.0\,\kev\
energy range.

\subsubsection{Swift Snapshots}
\label{sec_swift_red}

Finally, \swift\ (\citealt{SWIFT}) has also sporadically taken snapshot
observations of the Circinus field. We searched the \swift\ archive for pointed
observations with at least 1\,ks duration, such that reasonable flux estimates
might be obtained, and found 5 observations that met our criteria. Cleaned
event files were generated in the standard manner with \xrtpipeline, and
spectral products were extracted with \xselect. Source spectra were taken from
circular regions of radius $\sim$30'', and background spectra from larger,
adjacent regions free of contaminating point sources. Ancilliary responses were
generated with \xrtmkarf, and we use the latest redistribution matrices available
in the \swift\ calibration database (v13). The \swift\ spectra were only grouped
to have at least 5 counts per spectral bin, such that even the observations with
the lowest S/N (OBSIDs 00037273001 and 00037273004) had at least 10
spectral bins across the 0.3--10.0\,\kev\ energy range.

\begin{figure}
\hspace*{-0.5cm}
\epsscale{1.12}
\plotone{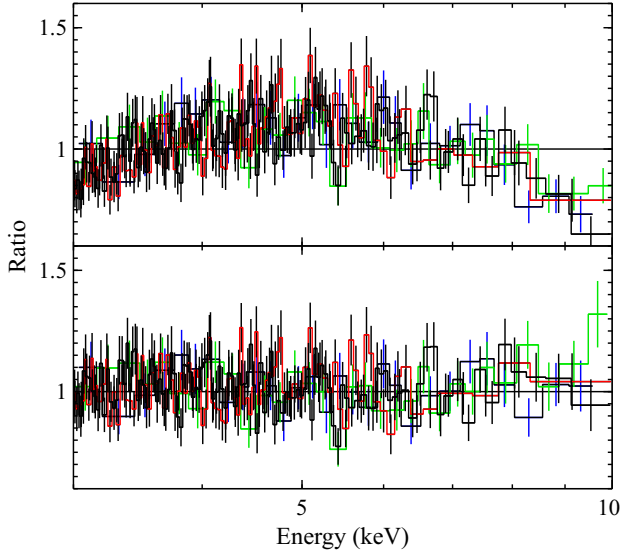}
\caption{
Data/model ratios for the simultaneous \xmm\ (black: \epicpn, red: \epicmos)
and \nustar\ (green: FPMA, blue: FPMB) datasets, modelled with both a powerlaw 
continuum modified by Galactic absorption (\textit{top panel}) and an unabsorbed
disk component (\textit{bottom panel}). The \xmm\ and \nustar\ datasets display
clear and consistent curvature across their common energy range (3--10\,\kev).}
\label{fig_cal}
\end{figure}

\section{Spectral Analysis}
\label{sec_spec}

The majority of our spectral analysis focuses on the long, higher S/N observations
of the ULX, \ie the joint 2013 \xmm\ and \nustar\ dataset, the 2006 \suzaku\
data and the 2001 \xmm\ data. Throughout this work, spectral modelling is
performed with XSPEC v12.8.0 (\citealt{XSPEC}), and quoted uncertainties on
spectral parameters are the 90 per cent confidence limits for a single parameter
of interest, unless stated otherwise. Unless stated otherwise, spectral fitting is
performed through \chisq\ minimisation. Neutral absorption is treated with
\tbnew\footnote{\tbnewlink}, the latest version of the \tbabs\ absorption code
(\citealt{tbabs}), with the appropriate solar abundances. Unless stated otherwise,
all models include Galactic absorption with a column of $N_{\rm H; Gal} = 5.58
\times 10^{21}$\,cm$^{-2}$ (\citealt{NH}). In the following, data from a variety
of X-ray missions are utilised, many of which operate multiple detectors
simultaneously (\eg \epicpn\ and \epicmos\ aboard \xmm). In these cases, the
data from the different detectors are modeled simultaneously, with all 
parameters tied between the spectra. However, we attempt to account for any
residual internal cross-calibration uncertainties between the detectors by
including a variable multiplicative cross-normalisation constant. This value is 
almost always found to be within $\sim$5 per cent of unity for all such missions,
with the only exception being the 2001 \xmm\ data, owing to the unfortunate
position of the source on the \epicpn\ detector as discussed previously (see
section \ref{sec_red_xmm2}).

\subsection{NuSTAR and XMM-Newton in 2013}

\subsubsection{Cross-Calibration}

We begin our analysis with the recent broadband \xmm+\nustar\ spectrum.
When modeling this joint dataset, we treat possible issues with flux
cross-calibration between \nustar\ and \xmm\ in the same way as we do
cross-calibration uncertainties between different detectors within a single
mission (see above). The individual \nustar\ and \xmm\ datasets have
substantial spectral overlap, both covering the 3--10\,\kev\ energy range, from
which cross-normalisation constants can easily be constrained. In order to
demonstrate the spectral consistency between \xmm\ and \nustar, we initially
focus on this energy range.

Applying a simple powerlaw model, modified by Galactic absorption results in a
poor fit, with \rchi\ = 796/565 and clear curvature present in the residuals for
both the \xmm\ and \nustar\ data (Fig. \ref{fig_cal}, \textit{upper panel}).
Inspection of the full 0.3--10.0\,\kev\ \xmm\ data suggests the overall neutral
column is most likely in excess of the Galactic column, closer to $N_{\rm H; tot}
\sim 10^{22}$\,\atpcm, but even allowing for a column of this order does not
fully remove the curvature in the 3--10\,\kev\ bandpass (a total column of
$N_{\rm H; tot} \sim 10^{22}$\,\atpcm\ still has a very limited effect above 3\
\kev). We therefore conclude that the 3--10\,\kev\ continuum of \culx\ is
intrinsically curved, similar to other bright ULXs (\citealt{Stobbart06,
Gladstone09, Walton4517}). If we instead model the 3--10\,\kev\ data with a
curved continuum, simply parameterising the data with an unabsorbed \diskbb\
component (\citealt{DISKBB}), we obtain an excellent fit with \rchi\ = 563/564
(Fig. \ref{fig_cal}, \textit{lower panel}). Allowing the \xmm\ and \nustar\
temperatures to vary independently does not improve the fit at all (\rchi\ =
563/563), and we obtain $T_{\rm XMM} = 2.01 \pm 0.04$\,\kev\ and $T_{\rm
NuSTAR} = 2.04 \pm 0.06$\,\kev. Clearly the 3--10\,\kev\ spectra obtained
with \xmm\ and \nustar\ are fully consistent. Furthermore, the \nustar\ and
\epicmos\ fluxes agree to within $\sim$15\,\%.

\begin{figure}
\hspace*{-0.3cm}
\epsscale{1.12}
\rotatebox{0}{
\plotone{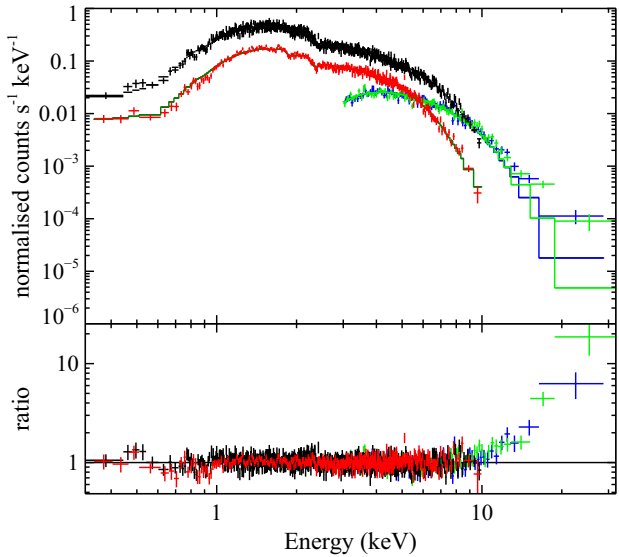}}
\caption{
The broadband 0.3--30.0\,\kev\ spectrum of \culx\ modelled with a simple
\diskbb\ accretion disk model. This model results in a clear excess in the
\nustar\ data above 10\,\kev.}
\label{fig_X2N2_diskbb}
\end{figure}

\pagebreak
\subsubsection{Continuum Modelling}

We now consider the full 0.3--30.0\,\kev\ broadband spectrum, and model the
\xmm\ and \nustar\ data simultaneously. In addition, we now (and hereafter)
formally includea both Galactic absorption and intrinsic neutral absorption (at the
redshift of Circinus), the latter being free to vary. Naturally, the simple powerlaw
continuum continues to provide a very poor fit (\rchi\ = 3388/1119). However,
applying the simple accretion disk continuum also results in a fairly poor fit
(\rchi\ = 1287/1119), and clear divergence between the data and the model can
be seen in the residuals at high energies ($\gtrsim$10\,\kev; see Fig.
\ref{fig_X2N2_diskbb}), where the Wien tail of the \diskbb\ model falls away far
faster than the data.

\begin{figure*}
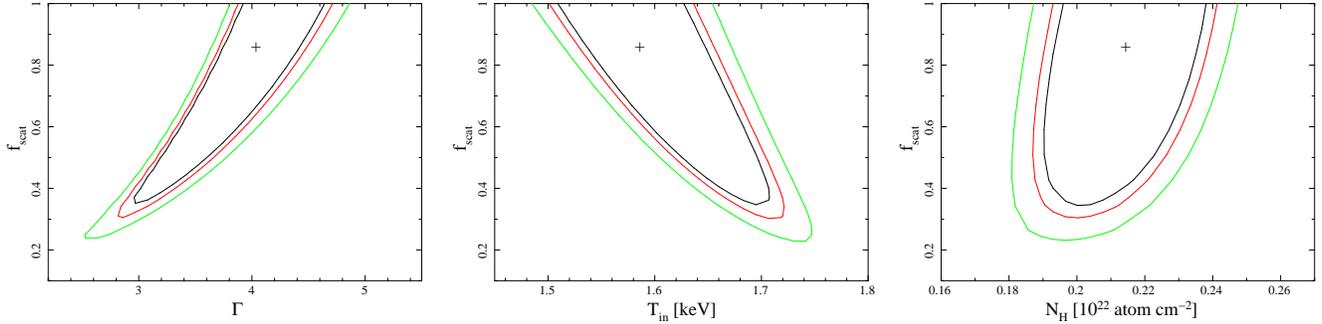

\epsscale{0.285}
\hspace*{-0.1cm}
\rotatebox{270}{
\plotone{./figs/contour_xmm2_nustar2_diskbb_simpl_Gam_Fsc.eps}}
\hspace*{-0.15cm}
\rotatebox{270}{
\plotone{./figs/contour_xmm2_nustar2_diskbb_simpl_Tin_Fsc.eps}}
\hspace*{-0.15cm}
\rotatebox{270}{
\plotone{./figs/contour_xmm2_nustar2_diskbb_simpl_Nh_Fsc.eps}}
\caption{
Two-dimensional \chisq\ confidence contours for various parameter
combinations from the \diskbb+\simpl\ model: the photon index ($\Gamma$),
disk temperature ($T_{\rm in}$) and the intrinsic neutral column ($N_{\rm H}$)
are each paired with the fraction of the diskbb flux scattered into the high energy
tail ($f_{\rm scat}$) in the \textit{left, centre} and \textit{right} panels
respectively.  In each case the black, red and green contours show the 90, 95, and
99\% confidence intervals (for 2 parameters of interest). Fairly strong parameter
degeneracies are observed in some cases.}
\label{fig_degen}
\end{figure*}

Initially, we attempt to model this additional high energy emission with a
powerlaw-like Comptonised component, applying the \simpl\ convolution model
(\citealt{simpl}), which `scatters' some fraction of an input seed photon
distribution into a high energy powerlaw tail, to the \diskbb\ continuum.
We use this model rather than a basic powerlaw component in order to ensure
that the powerlaw does not extrapolate to arbitrarily low energies, which is
potentially important given the high temperature of the disk ($\sim$2\,\kev). 
\simpl\ has three parameters, the photon index of the high energy tail, the
fraction of the seed flux scattered into the high energy tail ($f_{\rm scat}$), and
a flag determining whether to allow for both Compton up- and down-scattering,
or just the former. For simplicity, we only allow for up-scattering of the seed
photon spectrum, although the results obtained are not sensitive to this
assumption. The addition of this component significantly improves the fit, with
\rchi\ = 1137/1117, \ie an improvement of \dchisq\ = 150 for two additional
free parameters, and resolves the excess at high energies. The parameter values
obtained are quoted in Table \ref{tab_contin}. However, a number of parameters
are found to be degenerate with one another; this is particularly the case for the
parameters that determine the high energy spectrum (see Fig. \ref{fig_degen}).
Although fairly poorly constrained owing to these degeneracies, the photon
index obtained is very steep, $\Gamma = 4.0^{+0.4}_{-0.8}$.

The full 0.3--30.0\,\kev\ observed flux from \culx\ during this epoch is $(8.5
\pm 0.3) \times 10^{-12}$\,\ergpcmsqps. Owing to the steep high energy
spectrum, this is mostly dominated by the emission below 10\,\kev, which
contributes 85 per cent of the total 0.3--30.0\,\kev\ flux. At the distance of
Circinus ($D \sim 4$\,Mpc; \citealt{Freeman77, Koribalski04}), the broadband
flux corresponds to an extreme luminosity of $L_{\rm 0.3-30.0} = (1.63 \pm
0.06) \times 10^{40}$\,\ergps, assuming isotropic emission, placing \culx\
amongst the most luminous ULXs known to date, even before absorption 
corrections are considered. Although the exact correction is somewhat model
dependent, for the \diskbb+\simpl\ combination, the intrinsic 0.3--30.0\,\kev\
luminosity inferred is $\sim$2$\times 10^{40}$\,\ergps, a further $\sim$20 per
cent larger than the observed luminosity.

\begin{figure}
\hspace*{-0.3cm}
\epsscale{1.12}
\rotatebox{0}{
\plotone{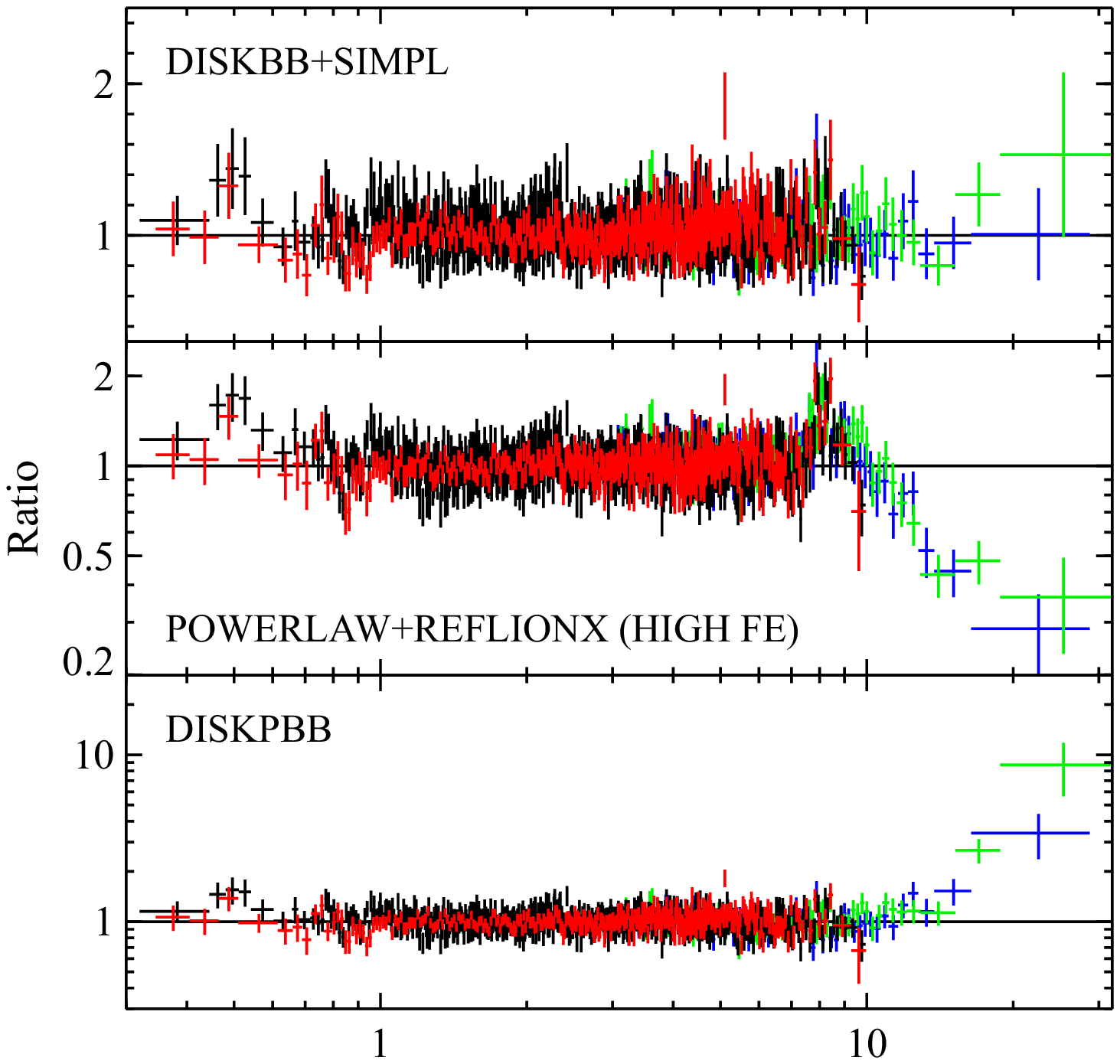}}
\caption{
Data/model ratios for a selection of the models applied to the combined
\xmm+\nustar\ dataset for \culx\ (see text for details). \textit{Top panel:} the
\diskbb+\simpl\ combination, which provides an excellent fit to the broadband
spectrum. \textit{Middle panel:} the relativistic disk reflection model in which
the 3--10\,\kev\ curvature is modelled as blurred iron emission/absorption (the
high iron abundance fit), which severely overpredicts the high energy \nustar\
data. \textit{Bottom panel:} the \diskpbb\ model, which still underpredicts the
high energy data, similar to the simpler \diskbb\ model.} 
\label{fig_X2N2_ratio}
\end{figure}

Given the steep nature of the high energy spectrum, it is not clear that the hard
excess can be considered to be similar to the hard excesses frequently seen in
AGN (\eg \citealt{Walton10hex, Walton13spin, Nardini11, Risaliti13nat,
Rivers13}) and BHBs (\eg \citealt{Zdziarski02, Corongiu03, Reis10lhs}), which
are best associated with Compton reflection (and if phenomenologically
modeled with a powerlaw would generally give $\Gamma<<2$). It has recently
been suggested that the combination of iron emission and absorption in a
relativistically smeared reflection spectrum from the inner regions of the
accretion disc might be able to explain the curvature observed below 10\,\kev\
in bright ULXs (\citealt{Caball10}). This would then allow the intrinsic high
energy emission to have a powerlaw-like form, as is typical for sub-Eddington
coronal emission. In general, this interpretation required high iron abundances
and strong relativistic broadening in order to reproduce the smooth
3--10\,\kev\ curvature (\citealt{Caball10, Walton4517}). However, while such a
model, consisting of a powerlaw-like corona and a smeared reflection component
(modelled with a  combination of the \reflionx\ reflection code, \citealt{reflion},
and the \relconv\ relativistic kernel, \citealt{relconv}) does provide an adequate
fit to the \xmm\ data alone (\rchi\ = 1018/950), when fit to the broadband
spectrum the Compton reflection hump at $\sim$20\,\kev\ is significantly in
excess of the observed \nustar\ data, as is clear from Fig. \ref{fig_X2N2_ratio}
(see also \citealt{Walton4517}), and the resulting fit is rather poor (\rchi\ =
1623/1111).

In this case, however, there is formally an alternative solution using this model
combination that provides an acceptable fit to the broadband spectrum (\rchi\
= 1145/1113), although it is rather different than previous applications to ULXs.
Rather than model the curvature with iron emission/absorption, this fit instead
attempts to remove all iron features, and requires the lowest iron abundance
allowed by the model.\footnote{The {\footnotesize REFLIONX} grid utilised is
calculated for photon indices in the range $\Gamma$ = 1.4--3.3, ionisation
states in the range $\log\xi$ = 0--4, and iron abundances in the range
0.1--10.0.} Without any iron absorption at $\sim$7\,\kev, the peak of the
Compton hump shifts to lower energies. In addition, the spin obtained is very
high, the radial emissivity index is maximized, the disk is required to be face on,
and its ionisation state is minimized. This combination serves to further reduce
the energy of the peak of the Compton hump, to the extent that the
3--10\,\kev\ curvature is actually modelled by this aspect of the reflected
emission. In fact, in this extreme corner of parameter space, the reflection
component is smeared and shifted to such an extent that, when absorbed by a
substantial neutral column, it takes on the appearance of a hot thermal-like
spectrum with a steep powerlaw tail. All the features typically associated with
reflected emission are essentially removed, which we interpret as further
evidence that the spectrum of \culx\ is not well modelled with traditional disc
reflection. Therefore, although statistically acceptable, we consider this to be a
very unsatisfying solution. If reflection is a relevant process for ULXs, the picture
must be more complex than the standard thin disk--corona accretion geometry.
Owing to the complexity of the model, we do not include the results obtained in
Table \ref{tab_contin}.

\begin{table*}
  \caption{Best fit parameters obtained for the variety of continuum models
  applied to the high S/N data available for \culx.}
\begin{center}
\begin{tabular}{c c c c c c c c c}
\hline
\hline
\\[-0.2cm]
Model & $N_{\rm H; int}$ & $T_{\rm in}$ & $p$ & $\Gamma$ & $f_{\rm scat}$ & $kT_{\rm e}$ & $\tau$ & \rchi\ (=$\chi^{2}$/DoF) \\
\\[-0.25cm]
& ($10^{21}$\,\atpcm) & (\kev) & & & (or $f_{\rm h}/f_{\rm s}$)\tmark[a] & (\kev) \\
\\[-0.25cm]
\hline
\hline
\\[-0.1cm]
\multicolumn{9}{c}{\xmm+\nustar\ (2013)} \\
\\[-0.15cm]
\diskbb\ & $1.6^{+0.2}_{-0.1}$ & $1.94 \pm 0.02$ & & & & & & 1287/1119 \\
\\[-0.25cm]
\diskpbb\ & $3.3\pm0.3$ & $2.22\pm0.06$ & $0.65\pm0.01$ & & & & & 1181/1118 \\
\\[-0.25cm]
\diskbb+\simpl\ & $2.1\pm0.2$ & $1.59^{+0.09}_{-0.04}$ & & $4.0^{+0.4}_{-0.8}$ & $>$0.43 & & & 1137/1117 \\
\\[-0.25cm]
\diskpbb+\simpl\ & $3.0^{+0.3}_{-0.7}$ & $2.0^{+0.1}_{-0.3}$ & $0.67^{+0.05}_{-0.02}$ & $<$3.9\tmark[b] & $0.08^{+0.55}_{-0.03}$ & & & 1131/1116 \\
\\[-0.25cm]
\diskbb+\comptt\ & $2.1\pm0.2$ & $1.5^{+0.3}_{-0.2}$ & & & $0.38^{+0.17}_{-0.30}$ & $>$2.8 & $<$20 & 1140/1114 \\
\\[-0.25cm]
\comptt$_{1}$+\comptt$_{2}$ & $1.0^{+0.8}_{-0.6}$ & $0.31^{+0.06}_{-0.05}$ & & & $0.57^{+0.22}_{-0.19}$ & 1: $1.3^{+0.4}_{-0.2}$ & 1: $>$12 & 1132/1112 \\
\\[-0.25cm]
& & & & & & 2: $3.1^{+1.2}_{-0.6}$ & 2: $>$5 & \\
\\[-0.1cm]
\multicolumn{9}{c}{\suzaku\ (2006)} \\
\\[-0.15cm]
\pl\ & $9.5\pm0.5$ & & & $2.46\pm0.03$ & & & & 1029/618 \\
\\[-0.25cm]
\diskbb\ & $1.2^{+0.3}_{-0.2}$ & $1.30\pm0.02$ & & & & & & 639/618 \\
\\[-0.25cm]
\diskpbb\ & $3.3^{+0.6}_{-0.7}$ & $1.47^{+0.07}_{-0.06}$ & $0.61\pm0.03$ & & & & & 606/617 \\
\\[-0.25cm]
\diskbb+\simpl\ & $2.0\pm0.4$ & $1.04^{+0.10}_{-0.08}$ & & $4.0^{+0.9}_{-1.9}$ & $>$0.17 & & & 590/616 \\
\\[-0.1cm]
\multicolumn{9}{c}{\xmm\ (2001)} \\
\\[-0.15cm]
\pl\ & $4.6\pm0.4$ & & & $2.18\pm0.04$ & & & & 587/580 \\
\\[-0.25cm]
\diskbb\ & $<0.2$ & $1.33^{+0.02}_{-0.03}$ & & & & & & 820/580 \\
\\[-0.25cm]
\diskpbb\ & $4.6\pm0.4$ & $4.9^{+0.8}_{-0.8}$ & $0.483\pm0.005$ & & & & & 576/579 \\
\\[-0.25cm]
\diskbb+\simpl\ & $2.0^{+0.7}_{-0.5}$ & $0.6^{+0.1}_{-0.2}$ & & $2.2^{+0.2}_{-0.3}$ & $>$0.55 & & & 569/578 \\
\\[-0.2cm]
\hline
\hline
\end{tabular}
\\[0.125cm]
$^{a}$ For models that do not include {\footnotesize SIMPL}, hard and soft
component fluxes ($f_{\rm h}$ and $f_{\rm s}$ respectively) are calculated
extrapolating the model components over the energy range 0.01-100\,\kev.
In the two {\footnotesize COMPTT} model, the higher temperature component
is the harder of the two. \\
$^{b}$ {\footnotesize SIMPL} does not allow for photon indices below $\sim$1.1.
\label{tab_contin}
\end{center}
\end{table*}

If ULXs do represent a population of sources accreting at very high- or
super-Eddington rates, the expected emission from the accretion disc may in
fact be substantially different from the simple \cite{Shakura73} thin disc profile
assumed in the \diskbb\ model. As the accretion rate increases towards
substantial Eddington fractions, the scale height of the disc is expected to
increase, and advection becomes an increasingly important process
(\citealt{Abram88}), resulting in shallower radial temperature profiles and hence
giving the appearence of a broader, less-peaked emission profile from the disc.
In order to investigate whether a broader disc profile could potentially account
for the additional hard emission relative to the simple \diskbb\ profile, we
attempted to model the broadband spectrum with a \diskpbb\ model, which
includes the index of the radial temperature profile ($p$) as an additional free
parameter (\citealt{diskpbb}). This does offer a substantial improvement over
the pure \diskbb\ model, with \rchi\ = 1181/1118, and the radial temperature
profile obtained is indeed shallower than expected for a thin disc (\ie $p <
0.75$). However, as shown in Fig. \ref{fig_X2N2_ratio} we again see an excess
of emission over the \diskpbb\ model, although it is slightly weaker than in the
\diskbb\ case, owing to the disc emission being able to extend to higher
energies while still reproducing the observed curvature in the 3--10\,\kev\ 
bandpass. An additional component is still required. In fact, this is the case for
\textit{any} model invoked to explain the curvature below 10\,\kev\ that falls
away above 10\,\kev\ with a thermal Wien spectrum, including more detailed
accretion disc models (\eg \kerrbb; \citealt{KERRBB}) and optically thick
Comptonisation by cool ($\sim$2\,\kev) electrons (\eg \citealt{Gladstone09}).

Adding a Comptonised component (\simpl) to the \diskpbb\ model again
provides a clear improvement to the fit, with \rchi\ = 1131/1116, \ie an
improvement of \dchisq\ = 50 for two additional free parameters over the pure
\diskpbb\ model. However, the improvement over the \diskbb+\simpl\
combination is very marginal, \dchisq\ = 6 for one additional free parameter,
and the additional model complexity serves to further exacerbate the parameter
degeneracies already present with the \diskbb+\simpl\ model (Fig.
\ref{fig_degen}). The best fit radial temperature profile for the disc is only
marginally constrained to be shallower than expected for the thin disc case ($p
= 0.67^{+0.05}_{-0.02}$), and in this instance we find that the photon index is
only loosely constrained at all ($\Gamma < 3.9$).

\begin{figure}
\hspace*{-0.5cm}
\epsscale{1.12}
\plotone{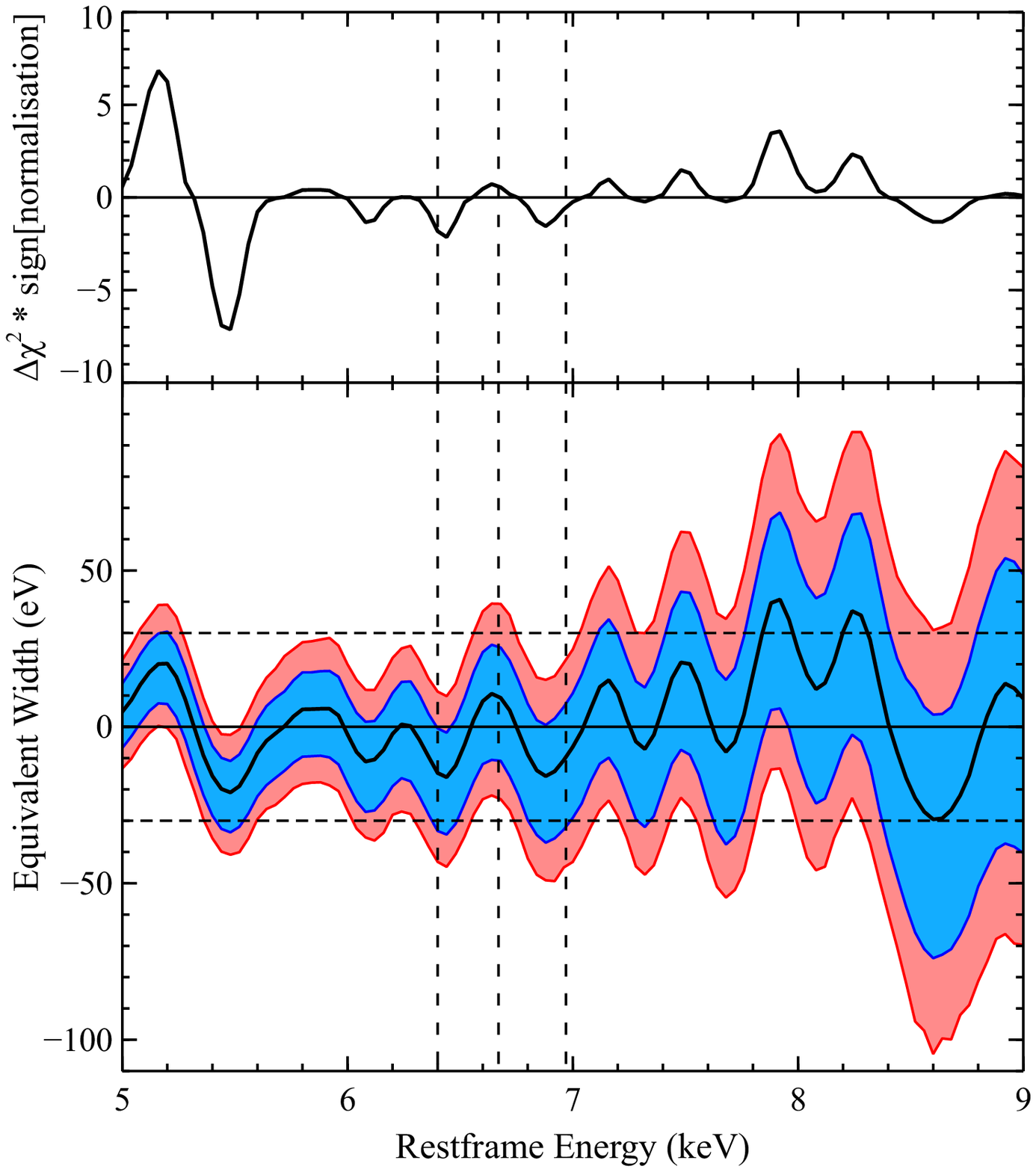}
\caption{
\textit{Top panel:} the $\Delta$\chisq\ improvement obtained with the addition
of a narrow Gaussian line, as a function of (rest frame) line energy, for the 2013
\xmm+\nustar\ dataset. Positive (negative) values of $\Delta$\chisq\ indicate
the  best fit line is in emission (absorption). We find no statistically significant
narrow iron K features. \textit{Bottom panel:} 90 (\textit{blue}) and 99\%
(\textit{red}) confidence contours for the equivalent width of the narrow line,
indicating the line strengths any undetected narrow features could yet have. For
clarity, the rest frame transitions of neutral, helium-like and hydrogen-like iron 
(6.4, 6.67 and 6.97\,\kev) are shown with vertical dashed lines. We also plot
dashed horizontal lines representing $EW=\pm30$\,eV, roughly indicative of the
absorption lines seen in the Galactic BHB GRS\,1915+105 (\citealt{Neilsen09}),
for comparison.
}
\label{fig_feKlimits}
\end{figure}

For completeness, we also fit the \diskbb+\comptt\ combination, frequently
used to parameterise the spectra from bright ULXs (\eg \citealt{Gladstone09,
Middleton11b, Walton4517, Walton12ulxFeK}), which allows for a variable
electron temperature for the thermal Comptonisation (\citealt{comptt}).
Unsurprisingly, this also provides an excellent fit (\rchi\ = 1140/1114),
although again the additional model complexity does not offer any substantial
improvement; the same model combination with the electron temperature fixed
at 500\,\kev, such that in the \nustar\ band the \comptt\ component is largely 
powerlaw-like, provides an equally good fit (\rchi = 1142/1115). There is again
substantial degeneracy between the various physical parameters, but despite
this there is a marked difference between the fit parameters for \culx\ and
those obtained for other bright ($L_{\rm X} \sim 10^{40}$\,\ergps) ULXs. Here,
it is the \diskbb\ component that primarily produces the 3--10\,\kev\
curvature, while in previous work this curvature is accounted for by the \comptt\
component, resulting in cool, optically thick electron distributions being inferred.
The \diskbb\ component instead usually accounts for the additional soft
emission seen below $\sim$1\,\kev\ in bright ULXs with less absorption (\eg
\citealt{Miller03, Miller04, Miller13ulx}). Unfortunately, owing to the fairly
substantial total absorbing column towards \culx, we are not highly sensitive to
the presence of any such emission. In this case, the \comptt\ component instead 
accounts for the excess emission observed above 10\,\kev. As such, the
Comptonisation parameters are not well constrained (see Table \ref{tab_contin}).
Nevertheless, the electron temperature obtained is still higher than typical results
from analyses limited to below 10\,\kev, which find $T_{\rm e} \sim 2$\,\kev, or
less (\eg \citealt{Gladstone09, Walton4517}).

Finally, we also consider a dual Comptonisation model for \culx, employing
two \comptt\ continuum components. Such dual-coronae have been proposed
for Galactic BHBs in some cases (\eg \citealt{Makishima08}), but this is the first
time such a model has been applied to a ULX. For simplicity, the seed photon
temperatures for the two components are linked throughout most of our
analysis. Initially, following \cite{Makishima08}, we attempted to fit the data
with a common electron temperature for each \comptt\ component, with the
two merely having differing optical depths. However, this resulted in a relatively
poor fit (\rchi\ = 1199/1113), with the model failing to correctly account for
the high energy emission, similar to Fig. \ref{fig_X2N2_diskbb}. Two different
electron temperatures are strongly required, as one of the components is
required to model the 3--10\,\kev\ curvature, while the other needs to extend
to higher energies in order to model the residual high energy excess. Allowing
for two different electron temperatures, an excellent fit is obtained (\rchi\ =
1132/1112). With this configuration, both electron distributions are found to
be optically-thick (see Table \ref{tab_contin}), although this is no longer the
case if the two components are allowed to have different seed photon
temperatures, in which case the parameters of the \comptt\ component
that accounts for the high energy excess are only poorly constrained, as before.

\subsubsection{The Iron K Region}
\label{sec_FeK}

The combined \xmm+\nustar\ dataset has sufficient photon statistics at high
energies to warrant an investigation of the iron K region (6--7\,\kev). Owing to
their typically moderate fluxes and their frequently soft spectra, ULX datasets
sensitive in the iron K energy range are naturally rare. For bright ($L_{\rm X} >
10^{40}$\,\ergps), isolated ULXs such studies have to date been limited to
Holmberg\,IX X-1 and NGC\,1313 X-1 (\citealt{Walton13hoIXfeK,
Walton12ulxFeK}).

To search for atomic features here, we follow the same approach undertaken in
\cite{Walton12ulxFeK, Walton13hoIXfeK}. We refer the reader to those
works for a detailed description of this approach, but in brief, we include a
narrow (intrinsic width of $\sigma = 10$\,eV) Gaussian, and vary its energy
across the 5--9\,\kev\ energy range in steps of 0.04\,\kev. The continuum
model used is the \diskbb+\simpl\ combination described above. For each line
energy, we record the $\Delta\chi^{2}$ improvement resulting from the inclusion
of the Gaussian line, as well as the best fit equivalent width ($EW$) and its 90 and
99\% confidence limits, calculated with the {\small EQWIDTH} command in \xspec,
using 10,000 parameter simulations based on the best fit model parameters and
their uncertainties.

The results obtained are shown in Fig. \ref{fig_feKlimits}; the top panel shows
the $\Delta\chi^{2}$ improvement, and the limits on $EW$ obtained are shown
in the bottom panel. For clarity, we highlight the energies of the \ka\ transitions
of neutral, helium-like and hydrogen-like iron, as well as $EW = \pm$30\,eV,
representative of the strongest iron absorption observed in GRS\,1915+105
(\citealt{Neilsen09}). As with our analysis of both Holmberg\,IX X-1 and
NGC\,1313 X-1, we find no statistically significant line detections. Any narrow
atomic features in the 2013 data in the immediate Fe K band (6--7\,\kev) must
have equivalent widths less than $\sim$50\,eV at 99\% confidence. The line
limits obtained here for \culx\ are not as stringent as those obtained most 
recently for Holmberg\,IX X-1 (\citealt{Walton13hoIXfeK}), but are similar to
those obtained previously for NGC\,1313 X-1 (\citealt{Walton12ulxFeK}).

\subsection{Suzaku in 2006}

\begin{figure}
\hspace*{-0.3cm}
\epsscale{1.12}
\rotatebox{0}{
\plotone{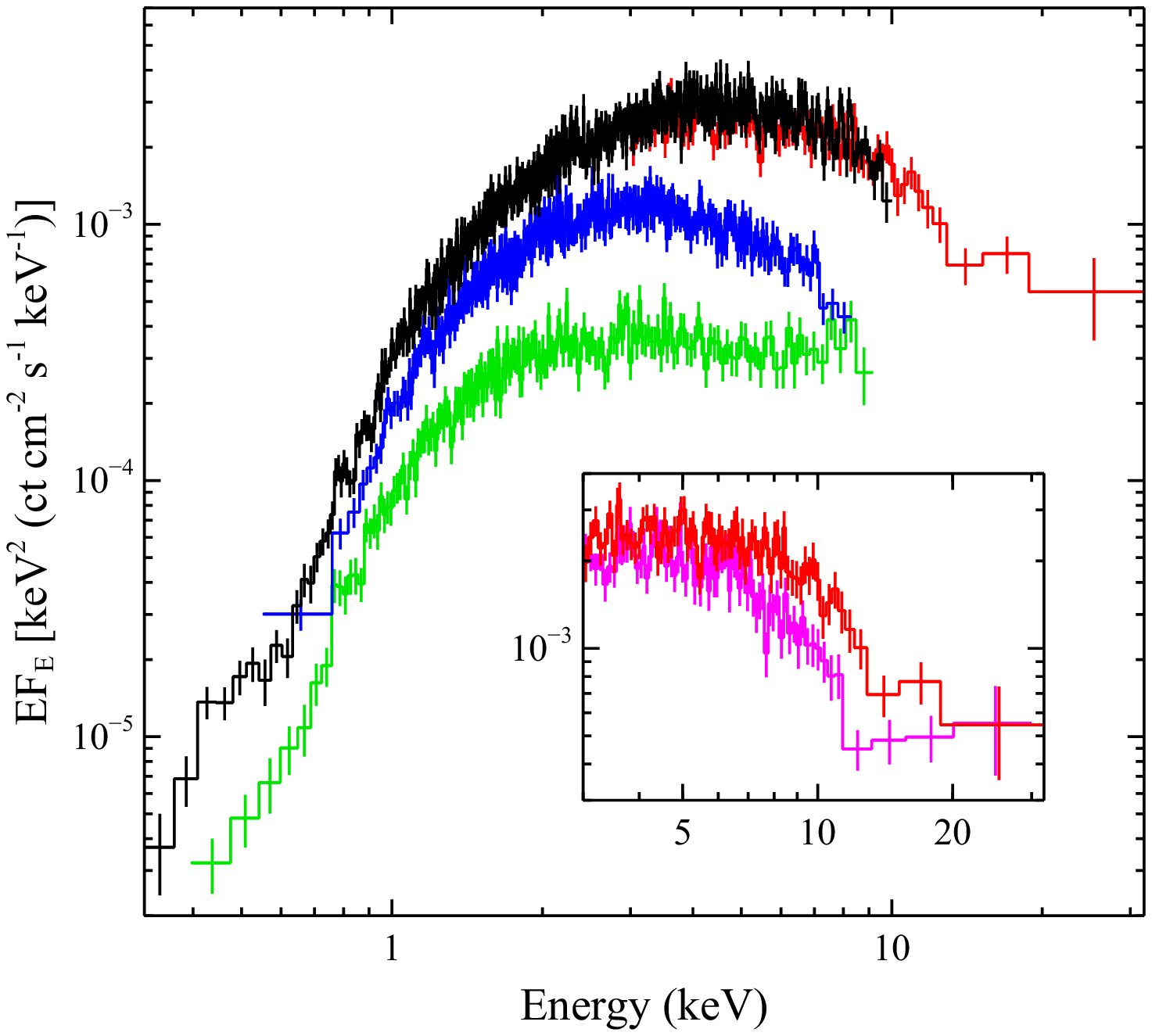}}
\caption{
The spectral evolution displayed by \culx. The 2013 \xmm\ (\epicpn) and
\nustar\ (FPMA) data are shown in black and red respectively, while the 2006
\suzaku\ (FI XIS) and the 2001 \xmm\ data are shown in blue and green. The
additional \nustar\ data are also shown in the inset in magenta, compared to the
same \nustar\ dataset shown in the main figure. All the data have been unfolded
through the same model, which simply consists of a constant.}
\label{fig_spec_comp}
\end{figure}

As with the more recent 2013 observations, the \suzaku\ spectrum obtained in
2006 is not well modelled by a simple absorbed powerlaw (\rchi\ = 1029/618),
requiring an intrinsically curved continuum instead. This is apparent from Fig.
\ref{fig_spec_comp}, in which the spectra from the three main epochs analysed
(2001, 2006, 2013) are directly compared, after having been unfolded through
the same simple model, consisting of just a constant. Indeed, the \suzaku\ data
have a distinctly thermal-like appearance. However, while a \diskbb\ continuum
gives a marked improvement and formally provides a statistically acceptable fit to
the data (\rchi\ = 639/618), an excess at high energies is again visible, broadly
similar to \xmm+\nustar\ dataset considered previously, albeit apparently
weaker and by necessity occuring at lower energies, owing to the limited 
bandpass of the XIS data.

In this case, the high energy excess can be resolved by allowing for a shallower
radial temperature profile for the disc with \diskpbb\ ($p = 0.61\pm0.03$),
and an excellent fit is obtained (\rchi\ = 606/617). However, this may be a
consequence of the limited bandpass, so we also consider a Comptonisation
origin for the high energy excess, again utilising the \diskbb+\simpl\
combination. Unsurprisingly an excellent fit is also obtained with this model
(\rchi\ = 590/616; see Fig. \ref{fig_XS_ratio}), but owing to the weak excess
and the lack of high energy data the \simpl\ parameters are again highly
degenerate, and therefore only poorly constrained individually. The observed
0.5--10.0\,\kev\ flux during this epoch, $(2.90 \pm 0.04) \times
10^{-12}$\,\ergpcmsqps, is significantly lower than observed in early 2013.
Given the quality of fit obtained, the parameter degeneracy already present with
this combination, and the lack of any obvious residuals, we do not consider the
more complex \diskpbb+\simpl, \diskbb+\comptt\ or \comptt+\comptt\
models here.

\subsection{XMM-Newton in 2001}
\label{sec_xmm1}

Again, for the early (2001) \xmm\ observation, we begin by modelling the data
with a simple absorbed powerlaw model. Remarkably, in contrast to the two
datasets considered so far, such a simple model actually provides an excellent
fit to this dataset (\rchi\ = 587/580). Again, the results obtained for the spectral
parameters are presented in Table \ref{tab_contin}. In contrast, the simple
accretion disk model provides a very poor fit (\rchi\ = 820/580), with the model
severely underpredicting the data above $\sim$5\,\kev. This indicates there is a
marked difference between this \xmm\ observation and the later \suzaku\ and
\xmm+\nustar\ datasets, in which the spectrum below 10\,\kev\ is generally
well modelled with thermal emission. The difference can clearly be seen in Fig.
\ref{fig_spec_comp}. While the \suzaku\ and \xmm+\nustar\ datasets display
curvature in the 3--10\,\kev\ bandpass, the 2001 \xmm\ dataset does not,
indeed appearing more consistent with a absorbed simple powerlaw-like
continuum. The observed 0.3--10.0\,\kev\ flux in this observation is $(1.93
\pm 0.04) \times 10^{-12}$\,\ergps, slightly lower again than the \suzaku\
dataset, and significantly lower than the \xmm+\nustar\ dataset.

\begin{figure}
\hspace*{-0.3cm}
\epsscale{1.12}
\rotatebox{0}{
\plotone{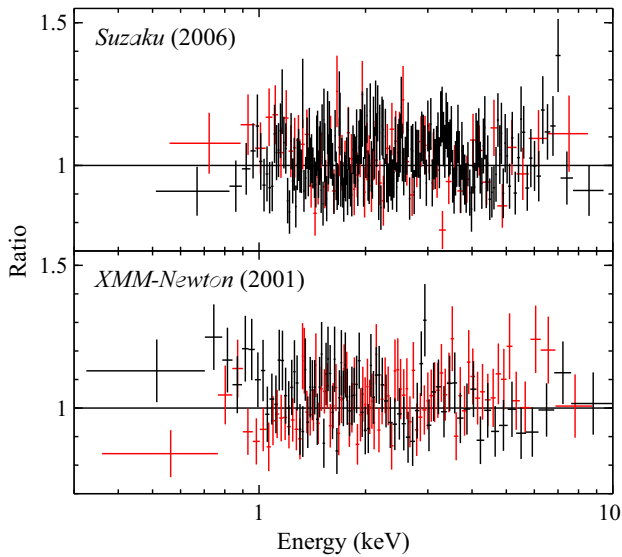}}
\caption{
Data/model ratios for the {\footnotesize DISKBB+SIMPL} model applied to the 
2006 \suzaku\ observation (\textit{top panel}) and the 2001 \xmm\ observation
(\textit{bottom panel}). Excellent fits are obtained in both cases (see Table
\ref{tab_contin}). Front-illiminatd XIS (\epicpn) and back-illuminated XIS (\epicmos)
data are shown in black and red respectively; the data have been rebinned for
visual clarity.}
\label{fig_XS_ratio}
\end{figure}

\begin{figure*}
\hspace*{-0.5cm}
\epsscale{1.12}
\plotone{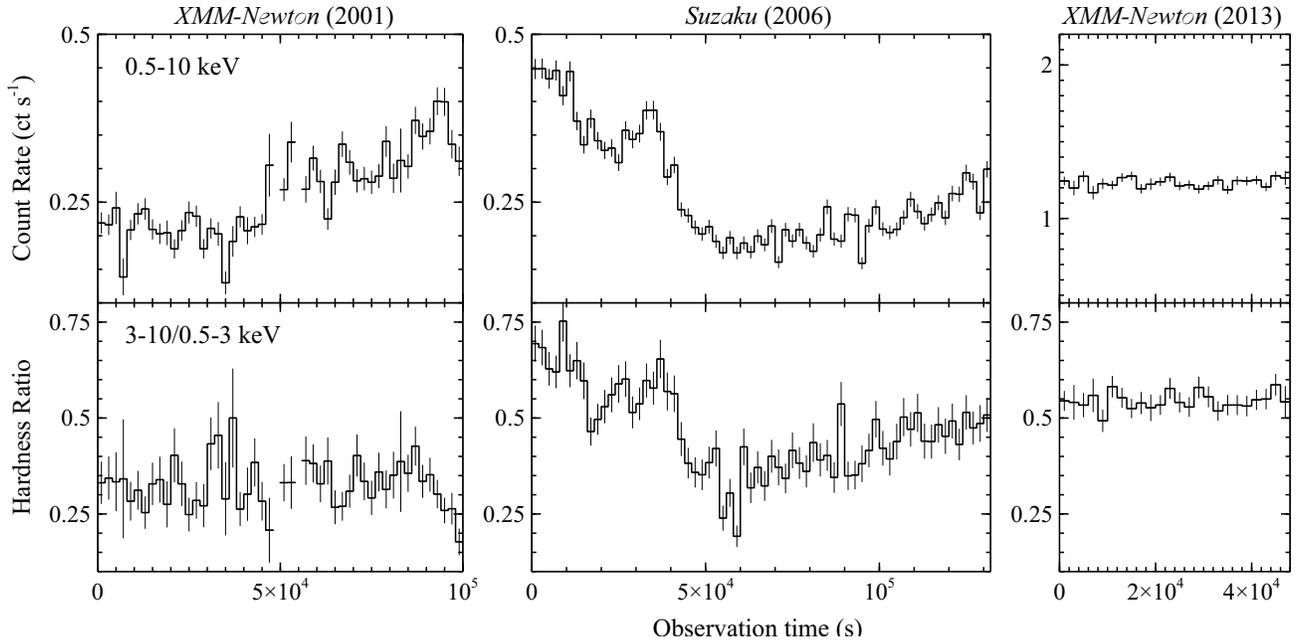}
\caption{
0.5--10.0\,\kev\ lightcurves (\textit{top panels}) and 3--10/0.5--3.0\,\kev\
hardness ratios (\textit{bottom panels}) for the 2001 \xmm\ (\textit{left panels}),
2006 \suzaku\ (\textit{centre panels}) and the 2013 \xmm\ observations
(\textit{right panels}). The axes in the top panels have been scaled to show a
similar dynamic range around the mean count rate for each observation. Clearly
contrasting short term behaviour is seen in each of these three observations (see
text).}
\label{fig_obs_lc}
\end{figure*}

For completeness, we apply some of the other models considered previously as
well. Statistically, the \diskpbb\ model offers a substantial improvement on the
simpler \diskbb\ model, however the parameters are pushed to truly extreme
values ($T_{\rm in} \simeq 5.0$\,\kev, $p < 0.5$) owing to the lack of curvature
in the 3--10\,\kev\ bandpass. Given the flux of this observation, such an
evolution of the disk would appear unphysical when compared to the more
moderate parameters obtained with the other datasets. We also again consider
the \diskbb+\simpl\ combination, in order to investigate the results obtained
interpreting the high energy ($\gtrsim$5\,\kev) excess observed with the
\diskbb\ model alone as Comptonisation. As shown in Fig. \ref{fig_XS_ratio},
this model gives an excellent fit (\rchi\ = 569/578), and actually provides a
reasonable improvement over the pure powerlaw continuum (\dchisq\ = 18 for
2 additional free parameters). However, the model is again dominated by the
powerlaw tail provided by the \simpl\ component, and we obtain a fully
consistent photon index to the pure powerlaw continuum. As with the \suzaku\
obervation, given the quality of fit obtained with the \diskbb+\simpl\ model,
and the lack of any obvious residuals, we do not consider the more complex
\diskpbb+\simpl, \diskbb+\comptt\ or  \comptt+\comptt\ models for this
dataset, and conclude that it is best respresented with a continuum dominated
by a powerlaw-like component with $\Gamma \sim 2$, perhaps with some mild
disc contribution at lower energies.

\begin{figure}
\hspace*{-0.5cm}
\epsscale{1.12}
\plotone{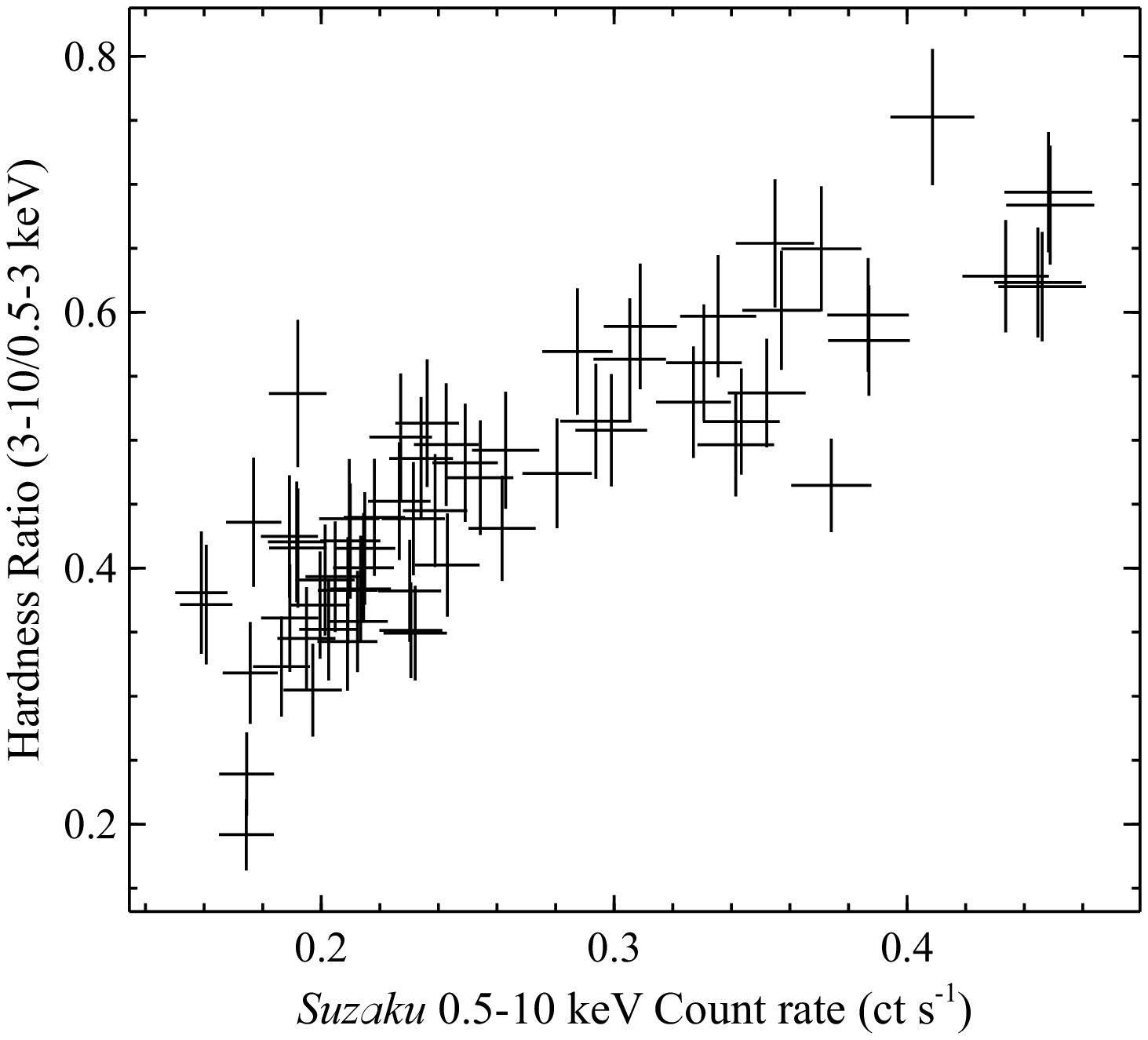}
\caption{
Hardness ratio--intensity diagram for the 2006 \suzaku\ observation. During this
epoch, the 3--10/0.5--3.0\,\kev\ hardness ratio clearly correlates with the full
0.5--10.0\,\kev\ count rate.}
\label{fig_hri}
\end{figure}

\section{Short Term Variability}
\label{sec_shortvar}

Figure \ref{fig_obs_lc} shows the 0.5--10.0\,\kev\ lightcurves for the three
longest duration observations, as well as the evolution of the
3--10/0.5--3.0\,\kev\ hardness ratio during these observations. Contrasting
short term behaviour can be seen from each of the three observations. The
long \xmm\ observation in 2001 shows clear flux variability, with no strong
associated spectral variability. The 2006 \suzaku\ observation also shows
strong variability, although in this instance there is clear spectral variability.
Indeed, this spectral variability appears to correlate extremely well with the
source flux, as shown in Fig. \ref{fig_hri}, with the source displaying harder
spectra at higher fluxes (note that \culx\ is vastly below the pile-up limit for
\suzaku). Finally, in 2013, \culx\ does not appear to show any strong short
term flux or spectral variability at all, although we note that of the three 
observations considered this has the shortest duration. In order to quantify
the differing levels of variability observed, we compute the fractional excess
variance (\fvar; \citealt{Edelson02, Vaughan03}) over the 0.5--10.0\,\kev\
energy range for each of the long  observations. For consistency, we divide the
earlier \xmm\ and \suzaku\ observations into 2 and 3 segments of
$\sim$45--50\,ks duration respectively, roughly that of the latest \xmm\
observation, and present the average value of \fvar\ obtained from these, in
order to ensure we are comparing the same timescales for each dataset. The
values obtained are presented in Table \ref{tab_fvar}, and confirm our earlier 
visual conclusions.

\begin{figure}
\hspace*{-0.5cm}
\epsscale{1.12}
\plotone{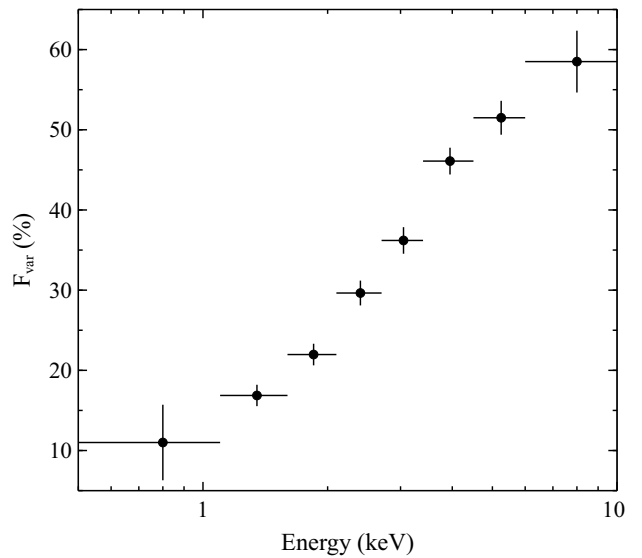}
\caption{
The fractional excess variance (\fvar) as a function of energy for the 2006
\suzaku\ observation. \fvar\ clearly increases with increasing energy.}
\label{fig_fvar_spec}
\end{figure}

\subsection{Spectral Variability}
\label{sec_LTs}

For the \suzaku\ observation, we also briefly investigate the nature of the 
observed spectral variability. First, we simply calculate the \fvar\ as a function
of energy. The resulting variability spectrum is shown in Fig. \ref{fig_fvar_spec}.
It is clear that the fractional variability increases monotonically with increasing
energy, and is strongest above the peak of the disk emission, at the energies at
which the powerlaw-like tail is most prominent in the \diskbb+\simpl\ model.
Second, we split the observation into seven segments, $\sim$15--20\,ks in
duration, and spectra are extracted for each following the data reduction
procedure outlined above (section \ref{sec_suz_red}). These seven segments are
modelled simultaneously with the \diskbb+\simpl\ combination, in order to
investigate the behaviour of the thermal component. As the \simpl\ parameters
were not well constrained when considering the full time averaged spectrum, we 
link the photon index between all the epochs in order to minimise the effects any
parameter degeneracies might have on the disk parameters obtained, which we
are primarily interested in. We also link the column densities of the intrinsic
absorption between the segments, as there is no strong evidence that this varies
here (see also \citealt{Miller13ulx}).

\begin{table}
  \caption{observed 0.5--10.0\,\kev\ fractional excess variability amplitudes.}
\begin{center}
\begin{tabular}{c c c}
\hline
\hline
\\[-0.25cm]
Mission & OBSID & 0.5--10.0\,\kev\ \fvar\ (\%)\\
\\[-0.3cm]
\hline
\hline
\\[-0.2cm]
\xmm\ & 0111240101 & $15\pm1$ \\
\\[-0.25cm]
\suzaku\ & 701036010 & $12\pm1$ \\
\\[-0.25cm]
\xmm\ & 0701981001 & $<2$ \\
\\[-0.25cm]
\hline
\hline
\\[-0.15cm]
\end{tabular}
\label{tab_fvar}
\end{center}
\end{table}

With this procedure, we obtain an excellent global fit to the seven segments,
with \rchi\ = 695/719. In addition to the disk temperature, which is a direct
product of the model, we also compute the intrinsic disk flux for each segment
with \cflux\ in \xspec. As shown in Fig. \ref{fig_LT}, there is a clear, positive
correlation between the inferred flux and the temperature of the disk
component. This evolution may also contribute to the energy dependence of the 
fractional variability. However, when modelled with a powerlaw relation, \ie $L 
\propto T_{\rm in}^{\alpha}$, accounting for the uncertainties on both the
temperatures and the fluxes with the algorithm described in \cite{Williams10},
the exponent obtained is much shallower than the expected $L \propto T^{4}$
relation for a standard thin accretion disc with a constant emitting area and a
constant color correction factor: $\alpha = 1.74 \pm 0.34$ (1$\sigma$ 
uncertainty).

\pagebreak
\section{Long Term Evolution}
\label{sec_longvar}

Given the observed correlation between the luminosity and temperature of the
disk component based on the short term variability during the 2006 \suzaku\ 
observation, we now wish to test whether this correlation also holds for the
long-term spectral evolution observed. Therefore, we take the same approach
outlined previously (section \ref{sec_LTs}) in order to investigate the evolution
of the disk parameters, but here making use of multi-epoch data. In addition to
the higher S/N datasets analysed previously, we also now consider the two other
pointed \nustar\ observations, obtained either side of the observation
coordinated with \xmm\ ({\small OBSIDs} 30002038002 and 30002038006).
The spectra from these two observations show some spectral and flux evolution
when compared to the \nustar\ data presented previously (see inset in Fig.
\ref{fig_spec_comp}), but are broadly consistent with one-another, and so we
combine them into a single dataset for the purposes of this analysis.

\begin{figure}
\hspace*{-0.5cm}
\epsscale{1.12}
\plotone{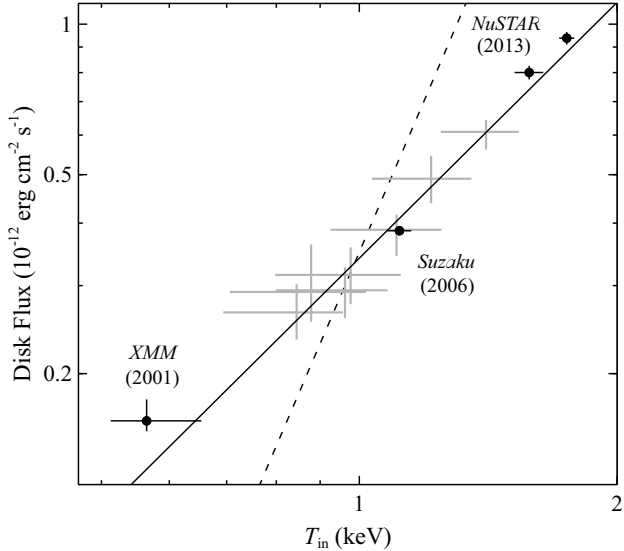}
\caption{
The luminosity--temperature relation inferred for the accretion disk of \culx\
when using the \diskbb+\simpl\ model. The multi-epoch data are shown in
black (see section \ref{sec_longvar}), while the individual \suzaku\ segments
considered are shown in grey (see section \ref{sec_LTs}). Remarkably, despite
probing very different timescales, all the data appear to follow a common
relation, which is significantly shallower than the naively expected $L \propto
T^{4}$ relation (shown as a dashed line) for standard stable disk emission. The
solid line shows the best fit to the multi-epoch data.}
\label{fig_LT}
\end{figure}

This approach also provides an excellent global fit to the average spectrum from
each epoch considered, with \rchi\ = 2486/2484. The common column density
obtained from all the datasets is $N_{\rm H;int} = 1.9^{+0.2}_{-0.1} \times
10^{21}$\,cm$^{-2}$, and the common photon index obtained is $\Gamma =
2.3^{+0.1}_{-0.2}$. Remarkably, as is clear from Fig. \ref{fig_LT}, we find that
despite the observations being taken over a span of more than a decade, the
multi-epoch evolution of the disk component is fully consistent with an
extrapolation of the short-term evolution observed from the \suzaku\ data
alone. Modeling the data again with a powerlaw relation, the multi-epoch 
exponent obtained is $\alpha = 1.70 \pm 0.17$, fully consistent with that
obtained previously.

\begin{figure*}
\hspace*{-0.5cm}
\epsscale{1.12}
\plotone{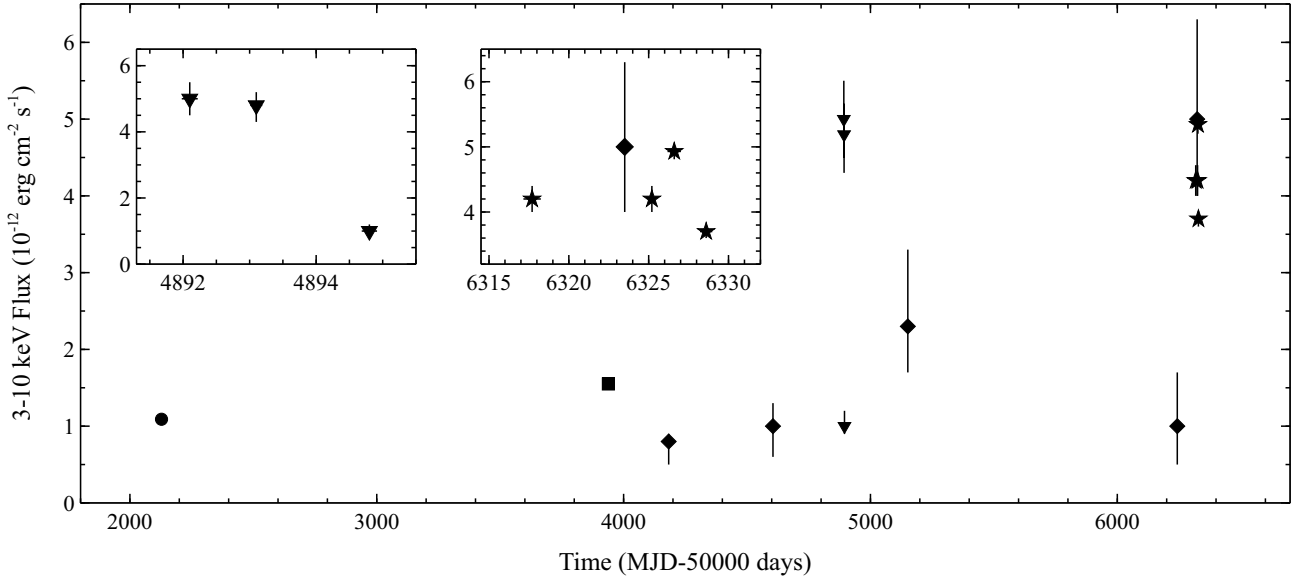}
\caption{
The long-term lightcurve for \culx, covering a period of over 10 years, compiled
from the new and archival data available. Periods with fairly high observing
cadence are shown in insets. Fluxes obtained with \xmm, \suzaku, \nustar,
\chandra\ and \swift\ are indicated with circles, squares, stars, triangles and
diamonds respectively.}
\label{fig_long_lc}
\end{figure*}

\subsection{Flux Evolution}
\label{sec_snapshot}

In addition to our analysis of the higher quality datasets available for \culx,
we also present a brief analysis of the lower S/N observations (the \swift\
snapshots, the short \chandra\ observations and the initial \nustar\
detection) in order to build up a long term lightcurve. Given the spectral
evolution apparent in Fig. \ref{fig_spec_comp}, we model each of these
datasets with a phenomenological powerlaw continuum with a variable
high-energy exponential cutoff, in order to allow for either curved or
powerlaw-like continua, as favoured by each individual dataset. The neutral
absorption is treated in the same manner as our more detailed spectral
analysis, including both a Galactic and an intrinsic absorption component.
Given the moderate quality (and/or the high energy nature) of the data, the
intrinsic column is fixed at $N_{\rm H;int} = 2.5 \times 10^{21}$\,\atpcm,
broadly consistent with the results obtained from the higher-quality
datasets. In the case of the \swift\ observations, spectral fitting is
performed through minimisation of the Cash statistic (\citealt{cstat}) owing
to the much less stringent rebinning applied to these data (see section
\ref{sec_swift_red}).

With this simple model, we compute the observed flux for these additional
exposures in the energy band common to all the missions utilised in this work,
3--10\,\kev. The fluxes obtained are quoted in Table \ref{tab_obs}, and the
long term lightcurve is shown in Fig. \ref{fig_long_lc}. Although the lightcurve
is sparsely sampled given the overall span of over a decade, strong long-term
variability is clearly apparent, with the 3--10\,\kev\ flux varying by at least
a factor of $\sim$5, and there are two clear periods of high flux observed in
early 2009 (\chandra\ datasets) and early 2013 (\nustar\ datasets). A more
comprehensive monitoring campaign on this source would be highly beneficial,
and would allow us to assess how frequently such high flux states occur.

\section{Discussion}
\label{sec_dis}

\subsection{Association With Circinus}
\label{sec_circinus}

Throughout this work, we have assumed that \culx\ is associated with the
Circinus galaxy. Here, we present a brief discussion of whether this is likely
to be the case, or whether \culx\ could plausibly be explained as distant
background active galaxy, or, particularly given the low Galactic latitude of
Circinus, as a foreground Galactic source.

\begin{figure*}
\hspace*{0.6cm}
\rotatebox{90}{
\epsscale{0.525}
\plotone{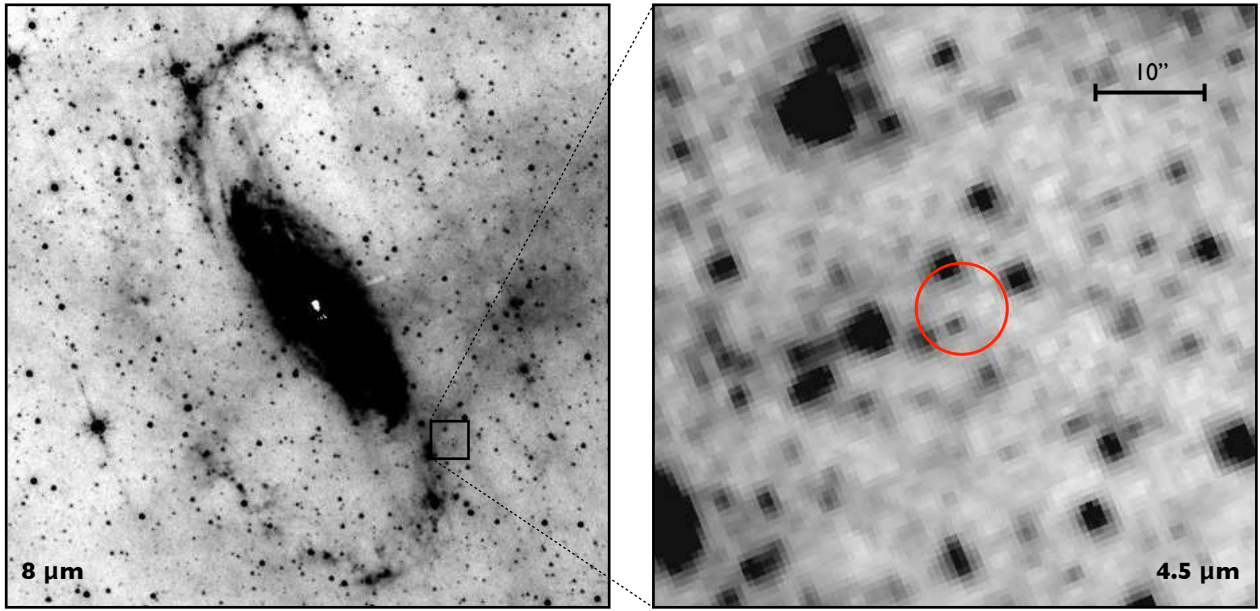}}
\caption{
\spitzer/IRAC imaging of \culx. \textit{Left panel:} wide-field IRAC 8 $\mu$m
(channel 4) image of Circinus, 14 arcmin on a side with North up and East to the
left, illustrating the location of ULX5 with respect to the spiral arms of the galaxy. 
\textit{Right panel:} Close-up IRAC 4.5 $\mu$m (channel 2), approximately 50
arcsec on a side, centered on ULX5. The red circle, with radius 4 arcsec, is
centered on the \nustar\ position of ULX5 (RA = $14^{h} 12^{m} 39^{s}$, DEC =
$-65$\deg $23' 34''$).
}
\label{fig_MIR}
\end{figure*}

First, we stress that the broadband X-ray spectrum of \culx\ obtained in 2013
is not consistent with that of a background AGN, which typically display
powerlaw spectra (\eg \citealt{Piconcelli03}). Unfortunately, at the time of
writing there is no Hubble coverage at the position of \culx\ with which to
perform a detailed search for optical counterparts which could assist in
classifying this source (\eg \citealt{Heida13}).  Instead, we have searched for
possible mid-infrared (MIR) counterparts in the wide-field \spitzer\
(\citealt{SPITZER}) Infrared Array Camera (IRAC; \citealt{SPITZER_IRAC}) map of
the Circinus galaxy obtained by \cite{For12}. Fig. \ref{fig_MIR} shows MIR images
of Circinus, with a wide-area $8 \mu$m view to highlight that \culx\ resides
near one of the spiral arms of this galaxy, and a $4.5 \mu$m zoom-in of the
X-ray position with a 4\arcsec\ radius circle illustrative of a conservative
estimate for the position uncertainty. There is one MIR source within this circle,
and an additional three MIR sources in close vicinity. All four sources are well
detected in channel 1 ($3.6 \mu$m) and channel 2 ($4.5 \mu$m) of IRAC, and
have relatively blue colors across this bandpass. In the Vega system,
[3.6]$-$[4.5] $\approx 0$ for all four sources, consistent with Galactic stars
and inconsistent with background AGN which typically have red MIR colors
(\eg \citealt{Stern05}). Therefore, we do not consider it likely that \culx\ is a
distant AGN being mis-identified as a ULX. Greater positional accuracy through a
dedicated on-axis \chandra\ observation will be required to determine which, if
any, of the \spitzer\ sources is the true NIR counterpart to \culx, and to aid in
future searches for optical counterparts.

The obvious candidates for Galactic sources that could also masquerade as a
bright ULX are foreground X-ray binaries (XRBs). As the Circinus galaxy is
roughly in the direction of the Galactic centre, the extent of the Galactic plane
towards \culx\ is $\sim$20\,kpc given our own location within the Galaxy
(\citealt{Sale10}). \cite{Jonker04} find that the typical scale-height of XRBs out of
the Galactic plane is roughly less than 1\,kpc, which contributes a negligible
amount to the maximum distance \culx\ could be at if within our Galaxy. Even at
the highest observed flux in 2013, \culx\ would therefore have a luminosity of
$L_{\rm X} \lesssim 3 \times 10^{35}$\,\ergps\ if Galactic, equivalent to $L_{\rm
X} / L_{\rm E} \lesssim 10^{-4}$ for a 10\,\msun\ black hole. Furthermore, these
observations revealed the source to have a very soft broadband spectrum.
Although Galactic black hole binaries (BHBs) are known to display soft spectra,
these are observed at high luminosities ($L_{\rm X} / L_{\rm E} \gtrsim 0.1$). In
contrast, low luminosity Galactic BHBs are generally observed to have hard
spectra (\eg \citealt{Remillard06rev}). Galactic neutron star XRBs can also display
soft broadband spectra, similar to that observed, but as with Galactic BHBs these 
are observed at high luminosities ($L_{\rm X} \gtrsim 5 \times 10^{36}$\,\ergps;
\citealt{Barret01rev}).

Finally, a Galactic cataclysmic variable (CV) would be consistent with the low
luminosity required to place the source in our own Galaxy. However, the
observed spectrum does not appear to be consistent with those of known CVs.
The high energy X-ray emission from CVs is generally observed to arise from a
multi-temperature collisionally ionised plasma, either from the boundary layer 
between the accretion disc and the white dwarf surface for non-magnetic CVs,
or from the post-shock plasma in the accretion columns for magnetic CVs; see
\cite{Mukai05} and \cite{Kuulkers06} for a recent reviews. However, as shown in
section \ref{sec_FeK}, the iron emission expected from such a plasma, which is
reliably observed in known CVs (see \citealt{Middleton12ip} for a particularly
extreme case), is not observed from \culx. Indeed, fits to the high energy
spectrum with thermal plasma models (\eg \citealt{raymondsmith}) with solar
abundances fail completely. Furthermore, the spectral evolution shown in Fig.
\ref{fig_spec_comp} is not typical behaviour for a Galactic CV.

Therefore, we conclude that none of these classes of Galactic X-ray source offers
an obvious observationally self-consistent scenario for \culx. A further point
worth highlighting, throughout all our spectral modelling we always require an
absorption column in excess of the Galactic column (as given in \citealt{NH}),
both when considering the local average for the column, and the measurement
closest to the source position, which also argues against a Galactic origin for
\culx. In combination with the arguments against a distant AGN origin presented
above, this strongly supports its association with Circinus. Finally, we note that
the observed variability rules out a young supernova remnant scenario for \culx,
leaving an extreme ULX (peak luminosity of $L_{\rm X} \sim 2 \times 10^{40}$
\ergps) as the only plausible interpretation. In addition, as discussed in the
following sections, \culx\ does display numerous similarities with other ULXs
that radiate at $L_{\rm X} \geq 10^{40}$\,\ergps, further supporting this
conclusion.

\subsection{Extreme Ultraluminous X-ray Sources}

Extreme ULXs with $L_{\rm X} \geq 10^{40}$\,\ergps\ are rare, with only a few
10s identified (\citealt{WaltonULXcat, Swartz04, Swartz11}). Furthermore, many
of these sources are reasonably distant ($D > 10$\,Mpc; \citealt{WaltonULXcat});
extreme ULXs close enough to enable detailed study are rarer still. In this work,
we have presented a multi-epoch spectral and temporal analysis of one of the
few such ULXs known, \culx, which  to date has received very little observational
attention, appearing only in the catalogue of \cite{Winter06}. This is likely in
part due to its location with respect to the Circinus galaxy, sitting outside the
D25 isophote in the relative outskirts of the galaxy, where the chances of an
observed source being foreground/background are fairly high. However, as
outlined previously, the  association of this source with the Circinus galaxy
appears to be robust. 

Observationally, \culx\ appears to show a number of similarities with other
extreme ULXs. First of all, the high quality datasets available when the source
was fairly bright show clear curvature in the 3--10\,\kev\ bandpass (see Fig.
\ref{fig_cal}). Such curvature is frequently observed in the spectra of other bright
ULXs (\eg \citealt{Stobbart06, Gladstone09, Walton4517, Walton13hoIXfeK};
Bachetti \etal\ 2013, \textit{submitted}), and does not seem to be consistent
with the powerlaw-like emission expected from a standard, optically thin
sub-Eddington corona. 

In addition, based on the compilation of serendipitous detections available, it is
clear that \culx\ can vary in flux from epoch to epoch by at least a factor of
$\sim$5 (see Fig. \ref{fig_long_lc} and Table \ref{tab_obs}). This, too, is broadly
similar to the level of long-term variability displayed by other ULXs. For example,
the recent \swift\ monitoring campaigns on Holmberg\,IX X-1 and NGC\,5907
ULX1 revealed long term variability by a factor of $\sim$3--4 (\citealt{Kong10,
Sutton13}), and multi-epoch \xmm\ observations of NGC\,1313 X-1 reveal long
term variability by a similar factor of $\sim$3 (\citealt{Feng06}). One potentially
subtle difference is that many of the brighter ULXs appear to show suppressed
short-term variability within single observations (\ie timescales less than
$\sim$100\,ks; \citealt{Heil09}). In this case, we were fortunate enough to
observe such variability from \culx, particularly during the 2006 \suzaku\
observation. However, it is noteworthy that during the most recent observation
of reasonable duration, when \culx\ was at its brightest, the short-term
variability also appeared to be suppressed, similar to the results presented by
\cite{Heil09}. Furthermore, high quality observations of ULXs with which
variability can properly be studied are relatively rare, so in many cases the
multi-epoch evolution of the short-term variability is not well constrained, and
may well be similar to that observed here.

\subsection{The Broadband X-ray Spectrum of \culx}

The \nustar\ observations of \culx, along with the observations of NGC\,1313
X-1, IC\,342 X-1 and Holmberg IX X-1 (see Bachetti \etal, \textit{submitted},
Rana \etal\ and Walton \etal, \textit{in prep}, respectively), represent one of the
first times it has been possible to reliably constrain the spectrum of an extreme
ULX above 10\,\kev. In combination with \xmm, we have been able to constrain
the broadband spectral form of \culx\ over the 0.3--30.0\,\kev\ bandpass
during a historically high flux state. During this epoch, \culx\ displayed
numerous observational similarities at lower energies ($\leq$10\,\kev) to
previous observations of other extreme ULXs, as outlined above.

The key spectral similarity is the curvature observed in the continuum over the
3--10\,\kev\ energy range. Based on the available data, until recently limited
to below 10\,\kev, a variety of interpretations for this curvature have been
proposed in the literature, including a high temperature accretion disk
(\citealt{Watarai01}), optically thick Comptonisation in a cool corona
(\citealt{Gladstone09}), and relativistic disk reflection (\citealt{Caball10}).
The former two predict that the curvature should continue to higher energies,
falling off with a thermal Wien tail, while the latter predicts a much stronger
high energy spectrum owing to the Compton reflection hump (see
\citealt{Walton4517}).

Broadly similar to the other ULXs observed by \nustar\ to date, the emission
above 10\,\kev\ from \culx\ is fairly weak compared to the emission below
10\,\kev\ during this epoch. The spectrum peaks at $\sim$4--5\,\kev, and
then falls away fairly steeply. However, when the broadband \xmm+\nustar\
spectrum of \culx\ is fitted with any purely thermal model with a Wien spectrum
at high energies, the \nustar\ data shows a clear excess over the model at high
energies (see Fig. \ref{fig_X2N2_diskbb}); pure thermal models can therefore be
rejected in this case. Nevertheless, the excess is not strong enough to be
explained as the Compton hump if the 3--10\,\kev\ curvature is due to
relativistic iron features from the inner accretion disk, so disk reflection does
not appear to offer a viable solution in this case either. Instead, this high energy
excess is well modelled simply as a powerlaw tail to the lower energy curved
spectrum. In physical terms, the broadband spectrum can be well explained as
a relatively hot accretion disk with an additional high energy Comptonised tail,
or as a Comptonised spectrum from an optically thick corona of electrons with
a dual temperature distribution. However, given the more straightforward
comparison with the established disk--corona Galactic BHB paradigm, and the
multi-epoch spectral and variability properties observed, we prefer the former.

\subsection{An Emerging/Re-Emerging Accretion Disk Scenario}

One of the most striking aspects of the available data for \culx\ is the clear
spectral variability (see Fig. \ref{fig_spec_comp}). During the 2001 \xmm\
observation, the source appears to display a powerlaw spectrum, with no
apparent curvature across the 3--10\,\kev\ energy range, while in the higher
luminosity observations obtained with \suzaku\ in 2006 and \xmm+\nustar\ in
2013 the spectrum seems to be dominated by a thermal component, displaying
the curvature across the 3--10\,\kev\ bandpass now typically associated with
bright ULXs. This rather dramatic spectral evolution is reminiscent of the
sub-Eddington state transitions observed in Galactic BHBs (\eg
\citealt{Remillard06rev, Done07rev, Fender12rev} for reviews). In particular, the
evolution from the 2001 \xmm\ data to the later thermal-like observations
(2006, 2013) seems to be comparable to the transition from the
hard/hard-intermediate state, in which the spectrum is largely dominated by the
Comptonised emission from the corona, to the soft state, in which the spectrum
is dominated by the thermal emission from the accretion disk.

Observation of spectral state transitions in ULXs have been claimed for a number
of individual sources in the literature (\eg \citealt{Feng06}). However, in many of
these cases, the claims relate to relatively subtle changes in the observed
spectrum, rather than true evidence for state transitions in the traditional sense
displayed by Galactic BHBs, particularly when one bears in mind that the same
state can be observed in the same source over a range of luminosities.
Nonetheless, there are some ULXs that show spectral evolution as strong as
observed in \culx. The most compelling case for sub-Eddington transitions is
also the most luminous ULX currently known, ESO\,243-49 HLX-1
(\citealt{Farrell09}), with strong spectral evolution that follows the characteristic
state-transition cycle displayed by Galactic BHBs in outburst (\citealt{Servillat11}).
However, in many respects ESO\,243-49 HLX-1 is rather unique, displaying
apparently periodic outburst cycles over which the flux varies by more than an
order of magnitude (\citealt{lasota11, Godet12}), and the peak luminosity
($L_{\rm X} \sim 10^{42}$\,\ergps) is vastly in excess of that reached by the
majority of ULXs, so we caution against drawing strong comparisons between 
ESO\,243-49 HLX-1 and the rest of the ULX population, including \culx.
However, amongst the more `standard' ULX population there have been a few
cases in which the spectral evolution appears similar to that presented here,
notably for the ULXs in IC\,342 (\citealt{Kubota01, Kubota02}).

\begin{figure}
\hspace*{-0.5cm}
\epsscale{1.12}
\plotone{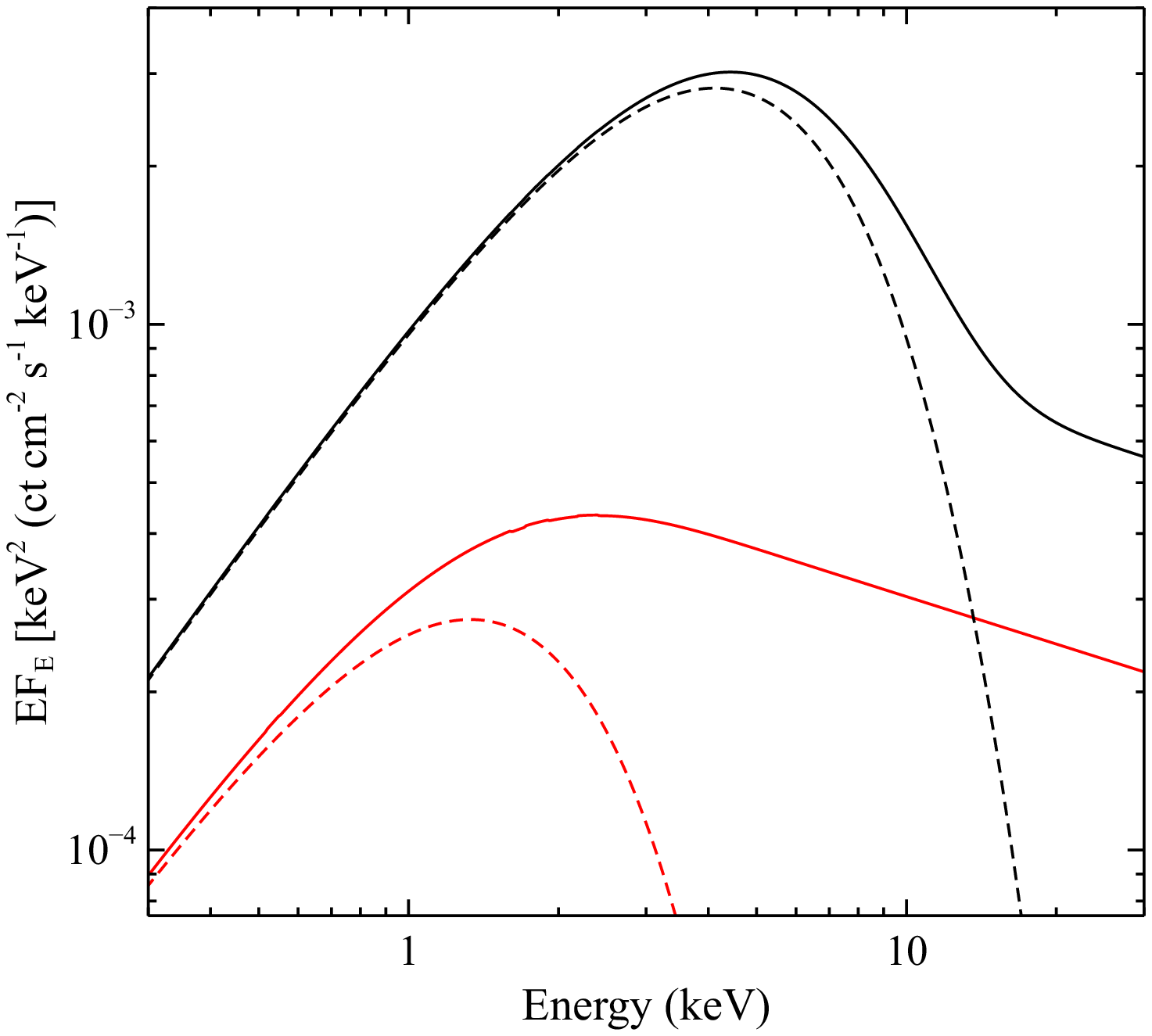}
\caption{
The total (unabsorbed) model (\textit{solid lines}) and the relative contribution
of the accretion disk (\textit{dashed lines}) for the 2013 \xmm+\nustar\
(\textit{black}) and 2001 \xmm\ (\textit{red}) observations, obtained with our
joint analysis of all the high S/N datasets with the {\footnotesize DISKBB+SIMPL}
model (see section \ref{sec_longvar}). The relative contribution of the accretion
disk over the analysed 0.3--30\,\kev\ energy range is much greater in the high
flux case.}
\label{fig_mod_evol}
\end{figure}

Indeed, although we have also considered more complex models, the high
quality spectra available from multiple epochs are all well modelled with a
simple combination of an accretion disc and a high energy Comptonised tail.
As the source becomes more luminous, the accretion disc becomes more
prominent with this combination, broadly similar to the behaviour observed
from Galactic BHBs. This is emphasized in Fig. \ref{fig_mod_evol}, which
shows both the total model and the relative \diskbb\ contribution for the
2013 \xmm+\nustar\ and the 2001 \xmm\ datasets (\ie the limiting flux
cases) obtained with our joint analysis of all the high S/N datasets with the
\diskbb+\simpl\ model combination (see section \ref{sec_longvar}). The
variability behaviour observed from \culx\ would also appear to support this
evolution, when considered in comparison to Galactic BHBs. Clear short-term
variability is observed in the two lower flux observations, in which the coronal
emission is most prominent. In particular, during the \suzaku\ observation,
we see that the fractional variability is stronger at higher energies (see Fig.
\ref{fig_fvar_spec}), where the coronal emission dominates. In contrast,
during the highest flux observation, in which the disk emission dominates
below 10\,\kev, no short-term variability is observed. In Galactic BHBs,
short-term variability is also associated with strong coronal emission (\eg
\citealt{Homan01, Churazov01}), with the harder states generally displaying
strong variability. In contrast, the thermal-dominated states display very little
variability, with the emission dominated by a relatively stable accretion disk.
The behaviour of \culx\ is strikingly similar.

However, identifying the spectral evolution in \culx\ with the hard-to-soft state
transition as seen in Galactic BHBs is not necessarily straightforward. In Galactic
sources, hard states (the canonical low/hard state and the hard-intermediate
state) can be observed at luminosities up to roughly 10--30\% of the Eddington
limit ($L_{\rm E}$). If we identify the luminosity observed in the 2001 \xmm\
observation with this Eddington ratio (\ie $L_{\rm X}/L_{\rm E} = 0.3$, to be
relatively conservative), the implied black hole mass is \mbh\ $\gtrsim
90$\,\msun. For the same spin, Eddington ratio and color correction factor, the
accretion disc temperature of such a black hole should be a factor of at least
$\sim$1.7 cooler than for a black hole of mass 10\,\msun. Accretion disks
observed from Galactic BHBs in the classical hard state are already rather cool,
with $T_{\rm in} \sim 0.2$\,\kev\ (\eg \citealt{Reis09, Reis10lhs, MReynolds13}).
However, the disk temperatures obtained here are actually rather similar to those
observed from such sub-Eddington Galactic BHBs at higher luminosities: $T_{\rm
in} \sim 0.5$\,\kev\ in the intermediate states and $\sim$1--2\,\kev\ in the
disk-dominated states (see \citealt{MReynolds13}).

Furthermore, although there is a clear positive correlation between the
temperature and inferred luminosity of the disk (see Fig. \ref{fig_LT}), which
remarkably seems to hold across all the timescales and luminosities
currently probed, the observed relation is $L \propto T_{\rm in}^{\alpha}$
with $\alpha = 1.78 \pm 0.19$, much shallower than the theoretically
expected relation for a standard, geometrically stable thin disk (\ie $L
\propto T^{4}$). While Galactic BHBs do themselves frequently show
significant deviations from this relation (\eg \citealt{Dunn11, MReynolds13}),
the strongest deviations tend to be seen either when the coronal emission
was strong (\ie hard states), probably linked to strong irradiation of the disk,
or very high luminosities, at which the scale height of the disk should start
to increase and advection becomes increasingly important. During the
thermal-dominated state, and while the disk remains thin, the observed
luminosity-temperature relation does tend to follow expectation relatively
well (see also \citealt{Gierlinski04LT}). At low luminosities, the deviations
away from $L \propto T^{4}$ tend to be in the sense that the disk
temperature is less dependent on the luminosity, becoming almost constant
(\citealt{MReynolds13}). In contrast, the \culx\ disk temperature displays a
stronger dependence on luminosity than expected.

This is instead more similar to the behaviour observed from another ULX,
NGC\,1313 X-2, based on the results obtained by \cite{Kajava09} modelling
\xmm\ and \chandra\ data with a pure \diskbb\ model ($\alpha = 2.39 \pm
0.16$). It is also similar to the deviations from $L \propto T^{4}$ displayed by
the Galactic BHBs GRO\,J1655--40 and XTE\,J1550--564 during the higher
luminosity stages of their respective outbursts, when the sources were in the
very-high state (\citealt{Kubota01gro1655, Kubota04, Saito06}). Sources in the
very-high state also display strong Comptonised emission, but with steeper
spectra and at substantially higher luminosities than the more traditional hard
states (\citealt{Remillard06rev}). However, it is interesting to note that, for the
latter cases, these deviations are still seen at sub-Eddington luminosities.
XTE\,J1650-500, another sub-Eddington Galactic BHB, also seems to display
similar behaviour, although these data are more limited (\citealt{Gierlinski04LT}).

The observed accretion disk evolution can depart from the naively expected
$L \propto T^{4}$ relation for a variety of reasons, which can be roughly
separated into changes in the disk geometry (\ie emitting area) and changes
in the detailed plasma physics of the disk atmosphere (\eg temperature
dependent opacities, evolving vertical structure and dissipation profiles,
\etc). The former primarily relates to changes in the inner radius, $R_{\rm
in}$, while the latter potentially incorporates a variety of complex effects
that are difficult to isolate observationally, and so the combined effect is
instead typically quantified as a multiplicative color-correction factor, that
relates the effective mid-plane temperature of the disk to that actually
observed: \Tin\ = \fcol\Teff. For simplicity, \fcol\ is usually assumed to be
energy independent, such that is only serves to shift the observed
temperature of the disc, rather than modify its spectral form. A substantial
body of work has been undertaken attempting to theoretically determine
the expected values of \fcol\ across a wide range of accretion regimes, \eg
\cite{Shimura95, Merloni00, Fabian04fcol, Davis05}, which typically suggest
that \fcol\ $\sim$ 1.7 for disk-dominated sub-Eddington accretion. Here we
are primarily interested in the relative rather than the absolute behaviour of
\fcol. The disk should evolve as $L \propto R_{\rm in}^{2}T_{\rm eff}^{4}$,
hence variation in either \rin\ or \fcol\ can result in the relation between the
luminosity and the observed temperature deviating from $L \propto T_{\rm
in}^{4}$. In order to recover the luminosity--temperature relation observed
for \culx, either the inner radius of the disk must decrease with increasing
luminosity as \rin\ $\propto L^{-0.6}$ (for a constant \fcol), or the
color-correction factor must increase with luminosity as \fcol\ $\propto
L^{0.3}$ (for a constant \rin), or some combination of these effects is
present (observationally, changes in \rin\ and \fcol\ are unfortunately highly
degenerate, particularly for the simple models employed here). Therefore,
given the inferred range in disk luminosity, either \rin\ decreased by up to
a factor of $\sim$2.9, or \fcol\ increased by up to a factor of $\sim$1.7
between the 2001 and 2013 observations.

If we invoke a variable \rin\ to explain the observed spectral evolution, the
implication is that the disk was truncated beyond the innermost stable circular
orbit at least during the lower flux observations. Substantial truncation of the
accretion disk is only really expected at very low accretion rates ($L/L_{\rm E}
\lesssim 10^{-2}$), in the low-luminosity regime of the hard state (\eg
\citealt{Tomsick09}). However, while the spectrum of \culx\ during the 2001
\xmm\ observation could be considered relatively hard, particularly in
comparison to the more recent observations, and can be modeled simply as
a powerlaw, it does not seem consistent with expectations for such a
low-Eddington regime, during which Galactic BHBs typically display very hard
spectra ($\Gamma \lesssim 1.7$). Instead, the photon index obtained would
suggest one of the higher luminosity hard state manifestations (if the source
was in this regime at all). While it could be debated whether a factor of
$\sim$3 change in inner radius is `substantial' truncation, studies into the
evolution of the relativistic iron line profiles observed from Galactic BHBs at
various stages of their outbursts seem to rule out changes in the inner radius
even of this magnitude as sources evolve from the higher luminosity hard
states to the soft state (\citealt{Reis1752, Walton12xrbAGN}), and a variable
\fcol\ has been proposed as an alternative explanation for the continuum
evolution seen at these luminosities (\eg \citealt{MReynolds13, Salvesen13}).
In any case, as stated previously, the observed behaviour seems to compare
far more favourably to high-luminosity observations of Galactic BHBs, where
disk truncation is not expected, but evolution in the vertical structure of the
disk is. Therefore, it seems likely that the deviation away from $L \propto
T_{\rm in}^{4}$ is driven by a variable \fcol\ in this case.

The identification of \culx\ as a (reasonably) high-Eddington black hole
binary system actually appears to provide a fairly self-consistent picture, at
least to first order, when considered in the context of the observed/expected
behaviour of Galactic BHBs. As discussed previously, the observed
luminosity--temperature relation appears extremely similar to that inferred
from known Galactic binaries at high luminosity, and seems to require a
color-correction factor that increases with increasing luminosity. Theoretical
consideration of the expected evolution of \fcol\ suggest that the opposite
trend should be seen at low luminosities (\citealt{Merloni00}), which appears
to be supported by observation (\citealt{MReynolds13, Salvesen13}).
Furthermore, the spectrum during the 2001 \xmm\ observation, which shows 
strong Comptonised emission, is remarkably similar to that observed from both
XTE\,J1550-564 and GRO\,J1655-40 at the transition (referred to by Kubota
\etal\ as the `anomolous regime') between the classic thermal state in which
the sources follow $L \propto T_{\rm in}^{4}$ rather well, implying a constant
\fcol, and the higher luminosities at which they appear to deviate from this
relation.

This is potentially of key importance, as it could provide an anchor from which
to estimate the relative luminosity of \culx, under the assumption that this
transition occurs over a fairly narrow range of $L/L_{\rm E}$. Although this is a
fairly strong assumption, as the behaviour of Galactic BHBs can be quite diverse,
XTE\,J1550-564 and GRO\,J1655-40 seem to make this transition at $L/L_{\rm
E} \sim 0.3$ and 0.1 respectively (\citealt{Gierlinski04LT}). Conservatively
adopting the former for the 2001 \xmm\ observation, we return to the previous
estimate of \mbh\ $\sim$ 90\,\msun. However, associating \culx\ with this
accretion regime, rather than the canonical low/hard state, has the advantage
that the \xmm\ disk temperature is indeed lower than the temperatures
observed from both XTE\,J1550-564 and GRO\,J1655-40 during the relevant
transition ($T_{\rm in} \sim 1$\,\kev), by roughly the factor expected for a
$\sim$90\,\msun\ black hole compared to a $\sim$10\,\msun\ black hole.
However, in the absence of any dynamical information on the putative binary
system, and given the diversity in the behaviour observed from Galactic BHB
accretion disks (\citealt{Gierlinski04LT, Dunn11, MReynolds13}), this mass
estimate must still be considered speculative, and treated with the appropriate
caution. Furthermore, we also urge caution in extrapolating the proposed 
identification of \culx\ to the general ULX population, as it is based on the
specific behaviour displayed by this source.

Such a black hole is just about at the upper limit of the mass range it is
currently believed possible to form \textit{in situ} via standard stellar evolution
(\citealt{Zampieri09, Belczynski10}). This would appear to require a low
metallicity, $Z \sim 0.05$\,\zsun\ or less. The work by \cite{Oliva99} does
suggest the Circinus galaxy has a sub-solar metallicity, although at roughly
$\sim$0.5\,\zsun\ this may not be low enough to form such a black hole
directly. If this mass is correct, more exotic formation mechanisms (\eg
\citealt{PortZwart04}) may be required. However, the metallicity estimates in
\cite{Oliva99} are based in circumnuclear clouds rather than the immediate
environment around \culx. Therefore, given that the mass is still ultimately
uncertain, we defer detailed discussion of possible formation scenarios until
more secure mass estimates and further studies focusing on the immediate
environment of \culx\ are available.

If our identification of the accretion regime displayed by \culx\ is correct, then
in the most recent high flux state we are observing the \textit{re-emergance} of
the accretion disk as it again begins to dominate over the very-high state corona,
and in principle at lower luminosities the X-ray spectrum should also appear
disk-dominated, as the source returns to the more traditional thermal state.
\cite{Tao13} suggest the very-high state corona could be related to the vertical
structure of the disk at high luminosities, and are able to reproduce similar
spectra to those observed with vertical dissipation profiles that dissipate more
energy at larger scale heights, in the hot, ionised upper layers of the disk
atmosphere. This disk state could also be described as having an \textit{energy
dependent} \fcol, or alternatively a distribution of \fcol\ values, that produces
the hard tail. This is potentially physically distinct from the corona that
dominates during the low/hard state, which may well be associated with the base
of a jet (\citealt{Markoff05, Miller12, Reis13corona}). 

In order to explain the observations of \culx\ as an extension of the high
luminosity behaviour of Galactic binaries, all of which are fully driven by an
evolving disk structure and without invoking very large mid-plane disk
temperatures, the evolution could roughly be along the following lines. As the
source increases in luminosity from the classic thermal state, the characteristic
height of the dissipation profile first increases relative to the scale height of the
disc, such that more energy is dissipated in the scattering atmosphere and a
strong, high energy tail emerges. Increasing the luminosity further still, the scale
height of the disc also increases and in effect catches up with the dissipation
profile, such that more of the energy is again dissipated in the optically-thick
regions of the disk, and the blackbody-like emission progressively dominates
again.

Obtaining lower flux observations may therefore provide a simple test of our
proposed identification. This would most likely require dedicated monitoring in
order to identify and  follow up periods of low flux. Although the temperature
of the disk should decrease further, the evolution should switch to roughly
following $L \propto T_{\rm in}^{4}$, which would alleviate the rate of decrease,
keeping the disk in the band observable with current soft X-ray instrumentation
for a wide range of luminosities.

\subsection{Super-Eddington Accretion and X-ray Outflows}

Galactic binaries in disk-dominated states frequently display evidence for
outflows in the form of narrow, highly-ionised iron absorption (\eg
\citealt{Miller06a, Neilsen09, Ponti12, King12}). Furthermore, in agreement
with basic expectation, the strength of these outflows appears to increase
with increasing luminosity (\citealt{Ponti12}). This prompted us to search for
evidence of similar features in the joint \xmm+\nustar\ dataset obtained in
2013, which offers both the best photon statistics in the iron \ka\ band and
the highest source luminosity of the higher quality datasets available (see
section \ref{sec_FeK}).

Similar to our analysis of other bright ULXs (\citealt{Walton12ulxFeK,
Walton13hoIXfeK}), we do not find any statistically compelling narrow iron
features in either absoption of emission, so in Fig. \ref{fig_feKlimits} we present
the equivalent width limits on any narrow lines that could have been present and
remain undetected. Across the immediate Fe K bandpass (6--7\,\kev), any
emission/absorption lines must have $EW \lesssim 50$\,eV. Although these
limits are not as stringent as recently obtained for \hoix\
(\citealt{Walton13hoIXfeK}), in absolute terms they still require any lines to be
weaker than the strongest features seen in Galactic BHBs (\citealt{King12}).
Metallicity is likely to be an issue here, with \cite{Oliva99} estimating the
(circumnuclear) iron abundance to be \feabund/solar $\sim 0.4$. However, even 
accounting for this iron abundance, the limits obtained still rule out the line
strengths that might be expected from simple scaling of the features in \eg
GRS\,1915+105 ($\sim$30\,eV; \citealt{Neilsen09}) up to the Eddington ratio
that would be inferred for a black hole of mass $\sim$10\,\msun\ ($EW \gtrsim
200$\,eV). Thus, it seems likely that we cannot be viewing the central regions of
\culx\ through any extreme super-Eddington outflow. Similar conclusions were
drawn for Holmberg IX X-1 and NGC\,1313 X-1. However, as noted earlier, we
again stress that the local metallicity is not well constrained, which if
substantially lower than the circumnuclear metallicity would further hinder line
detection. Nevertheless, a larger black hole accreting at a lower Eddington rate
also offers a plausible explanation for the lack of ionised absorption, as the
solid angle subtended by outflows launched from the accretion disk is widely
expected to increase with increasing Eddington luminosity (\eg \citealt{King09,
Dotan11, Kawashima12}), thus for a given observed luminosity there should be a
larger range of viewing angles that do not intercept any outflow launched for
larger black hole masses.

In addition, any iron emission must be weaker than observed from many Galactic
high mass X-ray binaries (HMXBs). Iron emission is ubiquitously observed from
such sources (\citealt{Torrejon10}), as they illuminate the strong stellar winds
launched by their massive binary companions. Following the discussion outlined
in \cite{Walton13hoIXfeK} for \hoix, we argue that the lack of strong iron
emission suggests that any stellar wind launched by the companion of \culx\ is
probably not sufficient to power the observed X-ray luminosities via wind-fed
accretion, and thus \culx\ most likely accretes via Roche-lobe overflow.

\section{Conclusions}
\label{sec_conc}

Prompted by a serendipitous detection with the \nustar\ observatory, we have
undertaken a multi-epoch spectral and temporal analysis of an extremely
luminous ULX located in the outskirts of the Circinus galaxy, utilising data from
most of the major X-ray observatories operating over the last decade, including
coordinated follow-up observations with \xmm\ and \nustar. Based on previous
detections of ULX candidates in Circinus, we refer to this source \culx. The
\nustar\ data presented here represent one of the first instances of a ULX reliably
detected at hard ($E > 10$\,\kev) X-rays. \culx\ is observed to vary on long
timescales by at least a factor of $\sim$5, and was caught in a historically bright
state by our 2013 observations, with an observed 0.3--30.0\,\kev\ luminosity of
$1.6 \times 10^{40}$\,\ergps. During this epoch, the source displayed a curved
3--10\,\kev\ spectrum, broadly similar to other bright ULXs. We consider a
variety of models for the broadband 0.3--30.0\,\kev\ spectrum obtained. Pure
thermal models (direct accretion disk emission, cool optically thick
Comptonization) result in a high energy excess in the \nustar\ data, and require
a second emission component. However, this excess is too weak for the
Compton reflection interpretation previously proposed for the 3--10\,\kev\
curvature in other ULXs.

In addition to the flux variability observed, \culx\ also displays strong spectral
variability from epoch to epoch, and even at times within a single epoch. All the
high quality datasets currently available are well modelled with a simple
combination of thermal accretion disk emission and a Comptonized corona, an
interpretation which is further supported by the observed short-term variability
properties. As the source luminosity increases, the accretion disk becomes more
prominent. However, although the disk temperature and luminosity follow a
common relation across all timescales probed, the observed relation is much
shallower than the $L \propto T^{4}$ relation naively expected for blackbody
radiation, varying instead as $L \propto T^{1.7}$. The spectral variability 
displayed by \culx\ is extremely reminiscent of that observed from the Galactic
BHBs XTE\,J1550-564 and GRO\,J1655-40 at high luminosities, which also seem
to roughly follow $L \propto T^{2}$. Identifying the lowest luminosity observation
of \culx\ with the transition into the $L \propto T^{2}$ regime, as the spectral
comparison would suggest, implies a black hole mass of $\sim$90\,\msun. This
is also consistent with the lower disk temperature displayed by \culx\ during this
epoch. However, we stress that given the fairly diverse behaviour observed from
Galactic BHB accretion disks, this mass estimate should be considered highly
uncertain. Further study of this remarkable source is certainly warranted in order
to see if this mass estimate truly holds up to scrutiny.

Finally, during the highest flux observation, we find no evidence for any iron
features in either emission or absorption, similar again to other bright ULXs.
Any features intrinsically present in the immediate Fe K bandpass must have
$EW\lesssim50$\,eV. The implication is that we are not viewing the central
regions of \culx\ through any extreme super-Eddington outflow, which would
also be consistent with \culx\ hosting a relatively massive black hole.

\section*{ACKNOWLEDGEMENTS}

The authors thank Koji Mukai for useful discussion regarding Galactic CVs, and
Rubens Reis for discussion regarding Galactic BHBs. This research has made use
of data obtained with the \nustar\ mission, a project led by the California
Institute of Technology (Caltech), managed by the Jet Propulsion Laboratory (JPL) 
and funded by NASA, \xmm, an ESA science mission with instruments and
contributions directly funded by ESA Member States and NASA, the \suzaku\
observatory, a collaborative mission between the space agencies of Japan (JAXA)
and the USA (NASA). In addition, this research has also made use of data
obtained from NASA’s \swift, \chandra\ and \spitzer\ satellites. We thank the
\nustar\ Operations, Software and Calibration teams for support with the
execution and analysis of these observations. This research has made use of the
\nustar\ Data Analysis Software (\nustardas) jointly developed by the ASI Science
Data Center (ASDC, Italy) and Caltech (USA). We also made use of the NASA/IPAC
Extragalactic Database (NED), which is operated by JPL, Caltech, under contract
with NASA. Some of the figures included in this work have been produced
with the Veusz plotting package: http://home.gna.org/veusz, written and
maintained by Jeremy Sanders. FEB acknowledges support from Basal-CATA 
(PFB-06/2007) and CONICYT-Chile (under grants FONDECYT 1101024 and Anillo
ACT1101). MB wishes to acknowledge the support from the Centre National
D\'Etudes Spatiales (CNES).

\bibliographystyle{/Users/dwalton/papers/mnras.bst}

\bibliography{/Users/dwalton/papers/references.bib}

\label{lastpage}

\end{document}